 \documentclass[final,5p,times,twocolumn,authoryear]{elsarticle}

\usepackage{amssymb}

\journal{Planetary \& Space Science}

\begin{document}

\begin{frontmatter}



\title{Modelling dust processing and the evolution of grain sizes in the ISM \\ using the method of moments}


\author{Lars Mattsson\corref{cor1}}

\address{Nordita, KTH Royal Institute of Technology \& Stockholm University, Roslagstullsbacken 23, SE-106 91 Stockholm, Sweden}
\ead{larsmat@kth.se}
\cortext[cor1]{Corresponding author. Tel: +46 8 553 784 23}

\begin{abstract}
Interstellar dust grains do not have a single well-defined origin. Stars are demonstrably dust producers, but also efficient destroyers of cosmic dust.  Dust destruction in the ISM is believed to be the result of SN shocks hitting the ambient ISM gas (and dust) and lead to an increased rate of ion sputtering, which reduces the dust mass. Grains located in cold molecular clouds can on the other hand grow by condensation, thus providing a replenishment mechanism or even a dominant channel of dust formation. In dense environments grains may coagulate and form large composite grains and aggregates and if grains collide with large enough energies they may be shattered, forming a range of smaller debris grains. The present paper presents a statistical modelling approach using the method of moments, which is computationally very inexpensive and may therefore be an attractive option when combining dust processing with, e.g., detailed simulations of interstellar gas dynamics. A solar-neighbourhood-like toy model of interstellar dust evolution is presented as an example. 
\end{abstract}


\end{frontmatter}


\section{Introduction}
\label{introduction}
Interstellar dust grains are abundant in a majority of gas-rich galaxies, including our own Milky Way, but the origin of these grains is complex. Stars are demonstrably dust producers, especially since it has become clear in recent years that some supernova remnants, without a doubt, contains very large amounts of dust \citep{Matsuura11,Gomez12a}, although the exact figures are uncertain \citep{Mattsson15a}. But stars are also efficient destroyers of cosmic dust. It is unclear to what extent the dust formed in supernovae (SNe) will actually survive the passage of the reverse shock and be mixed into the interstellar medium (ISM). Both theoretical and observational evidence suggest that much of the dust is in fact destroyed \citep{Bianchi07, Biscaro14,Lau15}. Dust grains are also consumed in star formation and can also be destroyed by shock waves from SNe. In the latter case destruction s believed to be the result of SN shocks hitting the gas and dust of the ambient ISM and thus lead to an increased rate of ion sputtering, which reduces the dust mass and changes the shape of the grain-size distribution \citep{McKee89,Draine90,Jones94,Slavin04,Bocchio14}. High star formation rates would consequently lead to high rates of dust destruction, which seems at odds with the existence of massive starburst galaxies with very large dust masses at high redshifts \citep{Bertoldi03, Beelen06, Gall11a,Gall11b,Mattsson11b, Michalowski10a,Michalowski10b,Watson15}. However, in cold molecular clouds in the ISM grains can also grow by condensation, thus providing a replenishment mechanism or even a dominant channel of dust formation \citep{Draine90,Mattsson11b,Kuo12}. Evidence from the radial distributions of dust in late-type galaxies suggest efficient grain growth by condensation in the ISM \citep{Mattsson12a,Mattsson12b,Mattsson14b} and the overall levels of dust depletion in damped Ly-$\alpha$ systems and quasar host galaxies point in the same direction \citep{DeCia13,DeCia15,Kuo12,Mattsson14a}.

In dense environments, e.g., the cores of molecular clouds, grains may coagulate due to a high rate of grain--grain interactions and form large composite grains and aggregates. Detections of large grains in the dense cores of molecular clouds may be evidence of  coagulation taking place \citep{Hirashita13} and large pre-solar grains have clearly an aggregative formation history \citep{Andersen07,Bernatowicz97}.  Moreover, if grains collide with large enough energies, the result can be partial or complete fragmentation of the grains, forming a range of smaller debris grains \citep{Jones96,Slavin04,Hirashita09,Hirashita10}. Coagulation and fragmentation do not directly affect the mass density of dust in the ISM much, although the change in total surface area of the grains may have a significant effect on the condensation rate \citep{Hirashita12}. The rate of grain--grain interactions is much affected by environmental factors, such as the small-scale dynamics of the ambient medium. A commonly suggested driver of coagulation as well as fragmentation is therefore turbulence in the ISM \citep{Hirashita09,Hirashita10}.

As demonstrated by \citet{Asano13a} and \citet{Remy-Ruyer14}, among others, there exist a critical metallicity (or actually density of relevant molecules) at which grain growth becomes efficient. The growth velocity depends on several time-dependent  parameters, most notably the availability of growth material, but it can  be shown that the evolution of the average grain size may be important too, which has not been taken into account in many previous theoretical studies of dust evolution in galaxies \citep[see, e.g.,][]{Calura08,Dwek98,Edmunds01,Pipino11,Valiante11,Zhukovska08}. \citet{Kuo12} showed that in case large grains are shattered so that small dust grains dominate, grain growth becomes efficient at a much earlier stage, which may have profound effects on the dust masses of high-$z$ galaxies. Overall, the interactions between the different types of grain processing mentioned above can be decisive for the average efficiency of dust condensation in the ISM and, obviously, also have a profound effect on the grain-size distribution. Numerical models of dust evolution including interstellar condensation, coagulation and fragmentations exist \citep[see][]{Asano13b,Asano14}, but due to the relatively wide range of sizes (three orders of magnitude in radii) and need for high resolution (small size bins) explicit treatment of the grain-size distribution quickly becomes computationally expensive. For this reason \citet{Hirashita15} suggested a simplified ``two-size approximation'', which largely reproduces the main features of full calculations. But in some cases more information about the grains is needed and other approaches to lowering the computational cost must be considered, even if some detailed information may still be lost. Note, however,  that information is lost due to binning errors occurring in direct numerical simulations as well.

In this paper, simplified equations of galactic dust evolution is combined with a statistical description of interstellar dust processing formulated using the method of moments. A scheme for numerical solution of the equations is discussed and a  simplistic model of the dust evolution in the solar neighbourhood is presented as an example (it should be emphasised that the purpose of this paper is not to model the dust component of the Galaxy, but to discuss a modelling method).

\section{Model and basic equations}
\label{model}
A model of interstellar dust processing and the evolution of the dust component must include a model of galaxy formation and evolution too. Here, a simple one-zone model, roughly corresponding to the solar neighbourhood, will be considered. Thus, a parametric standard model of Galactic chemical evolution is assumed. All basic parameters of the model are calibrated to match the observational constraints (see Table \ref{calibration}) on the mass assembly and chemical evolution of the solar neighbourhood as outlined in \citet{Mattsson10b}. To this model, equations describing the evolution of the grain-size distribution due to dust processing are added, with the aim of constructing an as computationally efficient model as possible.

\subsection{Galaxy Formation}
It is usually assumed that the Galaxy was formed through baryonic infall, or more precisely, by accretion of pristine gas (hydrogen and helium) so that the change of total baryonic mass density $\Sigma$ (the sum of the mass densities of stars and gas) at a given location in the Galaxy at a given time $t$ is equal to the infall rate $\dot{\Sigma}_{\rm inf.}(t)$, i.e., 
\begin{equation}
{d\Sigma\over dt} = {d\Sigma_{\rm stars}\over dt} + {d\Sigma_{\rm gas}\over dt} = \dot{\Sigma}_{\rm inf}(t)
\end{equation}
which also ensures mass conservation. Moreover, the baryonic mass of the Galaxy will here be divided into a halo/thick disc phase and a thin disc phase. The mass densities for gas and stars associated with these phases are related as follows,
\[
\Sigma_{\rm gas} = \Sigma_{\rm halo,\, gas} + \Sigma_{\rm disc,\, gas},
\]
\[
\Sigma_{\rm stars} = \Sigma_{\rm halo,\, stars} + \Sigma_{\rm disc,\, stars},
\]
\[
\Sigma_{\rm halo} = \Sigma_{\rm halo,\, gas} + \Sigma_{\rm halo,\, stars},
\]
\[
\Sigma_{\rm disc} = \Sigma_{\rm disc,\, gas} + \Sigma_{\rm disc,\, stars}.
\]
Thus, the total baryonic mass is $\Sigma = \Sigma_{\rm gas} + \Sigma_{\rm stars} = \Sigma_{\rm halo} + \Sigma_{\rm disc}$, which of course is needed in order to conserve mass.

It is often assumed that the rate of accretion follows an exponential decay \citep{Lacey83,Lacey85,Timmes95} and it is rather well established that the halo/thick disc and the thin disc components of the Galaxy were assembled on different time-scales and, perhaps, with some separation in time, as suggested by \citet{Chiappini97}. The latter scenario, known as the {\it two-infall model}, consists of two infall episodes where the disc formation starts after some time $\tau_{\rm max}$ (which is {when} the rate of accretion reaches its maximum), i.e., the total rate of accretion in the solar neighbourhood is
\begin{equation}
\dot{\Sigma}_{\rm inf}(t) = \left\{
\begin{array}{lll}
\dot{\Sigma}_{\rm halo}(t) & \mbox{if} & t \le \tau_{\rm max}\\
\dot{\Sigma}_{\rm halo}(t) + \dot{\Sigma}_{\rm disc}(r,t) & \mbox{if} & t > \tau_{\rm max}
\end{array}
\right.
\end{equation}
where dot notation is used to distinguish these prescribed rates from time derivatives being computed as part of the model (dot notation will be used for the star-formation rates as well). For the halo/thick disc, the rate of accretion is
\begin{equation}
\label{infall}
\dot{\Sigma}_{\rm halo}(t) = {\Sigma_{\rm halo}(t_0)\over\tau_{\rm halo}}
\left\{1-\exp\left[-{t_0\over \tau_{\rm halo}}\right]\right\}^{-1}\exp\left[-{t\over\tau_{\rm halo}}\right],
\end{equation}
and for the (thin) disc,
\begin{eqnarray}
\label{infall2}
\nonumber
\dot{\Sigma}_{\rm disc}(t) &=&
{\Sigma_{\rm disc}(t_0)\over\tau_{\rm disc}}\left\{1-\exp\left[-{(t_0-\tau_{\rm max})\over \tau_{\rm disc}}\right]\right\}^{-1}\times \\
&&\exp\left[-{(t-\tau_{\rm max})\over\tau_{\rm disc}}\right],
\end{eqnarray}
where $\tau_{\rm disc}$ is the infall time scale and $t=t_0$ is the age of the Galaxy. Here, the value used for $t_0$ is 13.5 Gyr, which can be regarded
as an upper limit.  For the halo/thick-disc phase it is assumed that $\tau_{\rm halo} = 1.0$ Gyr and for the thin disc  $\tau_{\rm disc} = 7.0$ Gyr.

The present-day total baryon density in the solar neighbourhood is assumed to be $51M_\odot$~pc$^2$
\cite[which is consistent with the results obtained by, e.g.,][]{McKee15,Holmberg04}.

\subsection{Star formation}
\begin{figure}
\resizebox{\hsize}{!}{\includegraphics[clip=true]{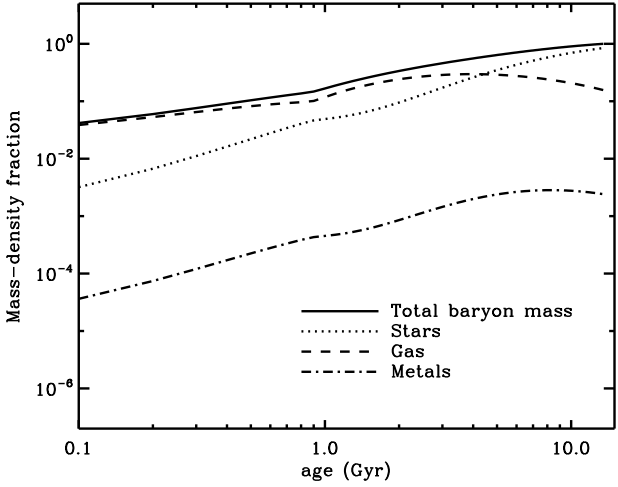}}
\caption{Time evolution of the baryon-mass components in the solar circle according to the adopted model.
\label{MW_sol_form}}
\end{figure}

During halo/thick--disc formation, the star-formation rate is prescribed by a modified Schmidt-law of the form
\begin{equation}
\label{sfr}
\dot{\Sigma}_\star(t) = \nu_{\rm halo}\,\Sigma(t)\,
\,\left[{\Sigma_{\rm gas}(t)\over \Sigma(t)} \right]^{1+\varepsilon},
\end{equation}
where $\nu_{\rm halo}$ is the star-formation efficiency expressed in Gyr$^{-1}$, $\varepsilon = 0.5$, $\Sigma = \Sigma_{\rm halo} + \Sigma_{\rm disc}$ and $\Sigma_{\rm gas} = \Sigma_{\rm halo,\,gas}$. This modified form gives a more burst-like star-formation history during the initial assembly of the halo/thick disc, which appears to fit the observational constraints slightly better. During thin-disc formation a regular Schmidt-law of the form
\begin{equation}
\label{sfr2}
\dot{\Sigma}_\star(t) = \nu_{\rm disc}\,\Sigma(t_0)\,
\,\left[{\Sigma_{\rm gas}(t)\over \Sigma(t_0)} \right]^{1+\varepsilon},
\end{equation}
is used. Note that $\Sigma$ is evaluated at  $t = t_0$, i.e., the present-day value is used, as opposed to Eq. (\ref{sfr}), where $\Sigma$ is a function of time. The parameterisation above is chosen for constancy of parameter units ($\nu_{\rm h}$ and $\nu_{\rm d}$ are both in Gyr$^{-1}$). A lower-threshold gas density for star formation is likely to exist \citep{Kennicutt89}, which suggests a need to modify also the Schmidt-law for the disc. However, since the gas density in the solar neighbourhood is not expected to fall short of this threshold and other parts the Galactic disc are not consider here, such a modification can be omitted in the present case. The time evolution of stars and gas for the parameters adopted here is shown in Fig. \ref{MW_sol_form}.

\subsection{The stellar initial mass function}
It is assumed that stars are formed according to a stellar IMF with a power-law tail. In its simplest form, the Galactic IMF is merely a power-law with sharp cut-offs at the lower and upper ends. The canonical power-law model first suggested by \citet{Salpeter55}, with a (negatively defined) logarithmic slope $x = 1.35$, has been revised as a consequence of more and better data. \citet{Scalo86} suggested $x = 1.7$ for the highest stellar masses and flatter trend for lower masses. In the last 20 years or so it has become evident that the IMF also smoothly turns over at low masses, so that the IMF has a well-defined most probable mass \citep{Chabrier03}. This also makes it a proper statistical distribution. Hence, an IMF of the form \citep{Larson98}
\begin{equation}
\label{imfunc}
\phi(m_\star)=\phi_0\,m_\star^{-(1+x)}\exp\left(-{m_{\rm c}\over m_\star} \right), \quad m_\star \le m_{\rm u},
\end{equation}
is adopted in the present paper, where $m_{\rm c}$, $m_{\rm u}$ are the masses defining the low-mass turn over and the high-mass truncation of the IMF, respectively. The high-mass slope is assumed to be $x = 1.7$ in accordance with \citet{Scalo86} and \citet{Chabrier03}. The constant $\phi_0$ is obtained by normalisation, i.e.,
\begin{equation}
\int_{0}^{\infty} m_\star\,\phi(m_\star)\,dm_\star = 1, 
\end{equation}
which yields
\begin{equation}
\phi_0 \approx {m_{\rm c}^{x-1}\over \Gamma(x-1)}, \quad x \ge 0, \quad m_{\rm u} \gg m_{\rm c},
\end{equation}
where $\Gamma(z)$ is the Gamma function. 

\subsection{Chemical evolution (enrichment of metals and dust)}
The fundamental equation describing the evolution of an element $i$ is
\begin{eqnarray}
\label{cee}
{d\Sigma_i\over d t} &=& \,\dot{\Sigma}_{{\rm inf},\,i}(t) - \frac{\Sigma_i(t)}{\Sigma_{\rm gas}(t)}\,\dot{\Sigma}_{\star}(t)+\\\nonumber
&&\,\int_{0}^{\infty} X_i(m_\star, t-\tau_{m_\star})\,\phi(m_\star)\,\dot{\Sigma}_{\star}(t-\tau_{m_\star})\,dm_\star,
\end{eqnarray}
where $X_i$ is the production matrix for an element $i$ \citep{Talbot71, Talbot73}, $\dot{\Sigma}_{{\rm inf},\,i}$ is the infall rate of $i$ and $\tau_{m_\star}$ is the {lifetime} of a star of mass $m_\star$. Additional terms may be added to include the nucleosynthetic contribution from type Ia supernovae (SNe Ia), but for simplicity such terms will be omitted here, since the focus of this paper is on the dust production [SNe Ia does not seem to show any signs of significant dust production either, see, e.g., \citet{Gomez09,Gomez12b}]. Most of the metals and dust is injected to the ISM by massive short-lived stars, which justifies use of the instantaneous recycling approximation (IRA) and the assumption of infall of pristine gas. That is, one may assume $\tau_m \ll \tau_Z$, where $\tau_Z$ is the timescale of metal enrichment, and $\dot{\Sigma}_{{\rm inf},\,i} = 0$, which means that eq. (\ref{cee}) can be simplified into
\begin{eqnarray}
\label{ira}
\nonumber
{d\Sigma_i\over d t} &=&  - \left[\frac{\Sigma_i(t)}{\Sigma_{\rm gas}(t)} - \int_{0}^{\infty} p_i(m_\star)\,\phi(m_\star)\,dm_\star\right] \dot{\Sigma}_{\star}(t)\\
&=&  \left[ y_{i} - \frac{\Sigma_i(t)}{\Sigma_{\rm gas}(t)}\right]\dot{\Sigma}_{\star}(t).
\end{eqnarray}
It should be noted that $d\Sigma_{\rm stars}/dt = (1-R)\,\dot{\Sigma}_\star$, where $R$ is the gas-return fraction (see below). The yield, defined as,
\begin{equation}
\label{yield}
y_i \equiv \int_{m_{\tau}}^{\infty} p_i(m_\star)\,m_\star\,\phi(m_\star)\,dm_\star,
\end{equation}
where $\phi(m_\star)$ is the mass-normalised IMF and $p_i$ is the typical fraction of the initial mass $m_\star$ of a star ejected in the form of an element $i$, may be treated as a constant. Stellar dust production is only increasing slowly with time once a certain metallicity is reached. Thus, a constant yield for stellar dust is also reasonable. 

Under the IRA, the value of the effective gas-return fraction $R$ for a generation of stars is obtained from the convolution of the IMF and the stellar remnant mass $w(m_\star)$, i.e.,
\begin{equation}
R = \int_{m_{\tau}}^{\infty} [m_\star-w(m_\star)]\,\phi(m_\star)\,dm_\star,
\end{equation}
where $m_\tau$ is the inversion of the stellar lifetime-mass relation evaluated at a suitable stellar lifetime which is still consistent with $\tau_m \ll \tau_Z$. The corresponding stellar mass is around $5\,M_\odot$. Using the \citet{Larson98} IMF with the parameters given in the previous section, the effective gas-return fraction is $R\approx 0.2$.  This value is used in all example models presented in Sect. \ref{results} 

Similarly, an equation for the dust can be written as
\begin{eqnarray}
\label{ira_dust}
\nonumber
{d\Sigma_{{\rm d},\,{j}}\over d t} &=& \left[ y_{{\rm d},\,{j}} - \frac{\Sigma_{{\rm d},\,{j}}(t)}{\Sigma_{\rm gas}(t)}\right]\dot{\Sigma}_{\star}(t) + \left({d\Sigma_{{\rm d},\,{j}}\over d t} \right)_{\rm dest} +\\
&&  \left({d\Sigma_{{\rm d},\,{j}}\over d t} \right)_{\rm cond} + \left({d\Sigma_{{\rm d},\,{j}}\over d t} \right)_{\rm coag} + \left({d\Sigma_{{\rm d},\,{j}}\over d t} \right)_{\rm frag}.
\end{eqnarray}
where $y_{{\rm d}, j}$ is the stellar yield of a dust species $j$. The additional terms, for destruction by sputtering, condensation, coagulation and fragmentation respectively, arises from expected interstellar processing of dust grains and will be described in detail the following sections.

  \begin{table}
  \small
  \begin{center}
  \caption{\label{calibration} Parameters and present-day ($t=t_0$) observational constraints of the Galaxy evolution model.}
  \begin{tabular}{cccc}
  \hline 
   Parameter  & Unit &  Mod. value & Obs. constraint \\
   \hline
   $\tau_{\rm halo}$ & Gyr & 1.0 & - \\
   $\tau_{\rm disc}$ & Gyr  & 7.0 & - \\
   $\tau_{\rm max}$ & Gyr  & 1.0 & - \\[1mm]
   $\nu_{\rm halo}$ & Gyr$^{-1}$ & 0.1 & - \\
   $\nu_{\rm disc}$ & Gyr$^{-1}$ & 0.37 & - \\[1mm]
   $h$ & kpc & 1 & 0.3 - 1.5 \\[1mm]
   $y_{Z}$ & - & 0.012 & 0.01 -- 0.015  \\[1mm]
   $\dot{\Sigma}_\star(t_0)$ & $M_\odot$~pc$^{-2}$~Gyr$^{-1}$ & $ 1.95 $ & 2 -- 5 \\
   $\dot{\Sigma}_{\rm inf}(t_0)$ & $M_\odot$~pc$^{-2}$~Gyr$^{-1}$ & $1.25$ & 0.5 -- 5 \\[1mm]
   $\Sigma_{\rm total}(t_0)$ & $M_\odot$~pc$^{-2}$ & 51 & 45 -- 56 \\
   $\Sigma_{\rm stars}(t_0)$ & $M_\odot$~pc$^{-2}$ & 43 &  35 -- 48\\
   $\Sigma_{\rm gas}(t_0)$ & $M_\odot$~pc$^{-2}$ & 8 & 7 -- 13 \\
  \hline
  \end{tabular}
  \end{center}
  \end{table}

\subsection{Production and processing of dust}
\subsubsection{Method of moments}
\label{moms}
It will be assumed that all grains are spherical, with a radius $a$ and thus a mass $m= 4\pi/3\,\rho_{\rm gr} a^3$. The grain-mass distribution (GMD) $f(m,t)$ is defined as the number (density) of grains $n_{\rm d}$ with masses on the interval $[m, m+dm]$, i.e. $dn_{rm d} = f(m,t)\,dm$. Properly normalised, $f$ could be regarded as a probability density function (PDF). The statistical (raw) moments of order $n$ of the grain-mass distribution $f(m,t)$ is then given by
\begin{equation}
\label{Mmom}
\mathcal{M}_n(t) \equiv \int_0^\infty m^n\,f(m,t)\,dm.
\end{equation}
Similarly, the moments of order $\ell$ of the grain-size distribution (GSD) $f(a,t)$, where $a$ is the grain radius, are given by
\begin{equation}
\label{Kmom}
\mathcal{K}_\ell(t) \equiv \int_0^\infty a^\ell\,f(a,t)\,da.
\end{equation}
In the following $f$ will be regarded as a function describing the number density of grains of a certain mass/size. The moments of $f$ are in such case related to physical quantities as
\begin{equation}
n_{\rm d} = \mathcal{M}_0 = \mathcal{K}_0, \quad \rho_{\rm d} = {\Sigma_{\rm d}\over h} = \mathcal{M}_1 = {4\pi\,\rho_{\rm gr}\over 3}\mathcal{K}_3,
\end{equation}
where $h = 1\,$kpc is the disc scale height.

The GSD $f$ must satisfy a ``continuity equation'' of the form
\begin{equation}
\label{eoc}
{df\over dt} \equiv {\partial f\over \partial t}+ \xi {\partial f\over \partial a} = \mathcal{S}(a,t),
\end{equation}
where $\mathcal{S}(a,t)$ represents sources and sinks, i.e, injection rate of stellar dust, dust destruction by sputtering in SN shocks and coagulation (or fragmentation) in the ISM, and $\xi$ is the rate of change of the grain radius due to processes which involve phase transitions, such as condensation and sublimation. In case of spherical grains, $\xi$ is size independent for condensation and sublimation, given that curvature effects can be omitted. The formal solution to this equation is
\begin{equation}
\label{conteq}
f(a,t) = f_0\left[a-\int_0^t\xi(t')\,dt'\right]+\int_0^t\mathcal{S}(a,t')\,dt',
\end{equation}
where $f_0$ is the GSD at $t=0$. Combination of eqns. (\ref{Kmom}) and (\ref{eoc}) yields 
\begin{equation}
{d\mathcal{K}_\ell\over dt} =   \ell\xi \int_{0}^{\infty} a^{\ell-1} f(a,t)\,da  + \int_{0}^{\infty} a^\ell \mathcal{S}(a,t)\,da. 
\end{equation}
Hence,
\begin{equation}
\label{momeq}
{d\mathcal{K}_\ell\over dt} = \ell\xi(t)\,\mathcal{K}_{\ell-1}(t)  + [{\rm sources }]_\ell - [{\rm sinks}]_\ell,
\end{equation}
where the first term on the right-hand side of the equation describes phase transitions (condensation and sublimation, but not sputtering induced by SN shocks) in terms of the rate of change $\xi$ of the grain radius. This is the general form of the moment equations. To formulate a self-consistent mathematical model of dust evolution, the adequate additional sources and sinks need to be specified, however. 

\subsubsection{Stellar production and consumption of dust}
Nucleation of dust grains is not expected to occur in the ISM since supersaturation is required and the partial pressures in the ISM are never that high. New grains must come from stars and the stellar dust-injection rate therefore replaces the nucleation rate normally introduced as a source term when modelling dust formation \citep[see, e.g,][and references therein]{Gail88}. This injection rate is given by
\begin{eqnarray}
\label{jstar}
\nonumber
\left({d\mathcal{K}_\ell\over dt} \right)_\star &=& {1\over \varsigma_\star}{K_\ell^\star(t-\tau_m)\over K_3^\star(t-\tau_m)} \int_0^\infty \,X_{{\rm d},\,j}(m_\star,t-\tau_m)\,\phi(m_\star)\times \\
&& \dot{\Sigma}_{\star}(t-\tau_m)\,dm_\star,
\end{eqnarray}
where $\varsigma_\star = 4\pi/3\,\rho_{\rm gr}\,h$, $K_\ell^\star$ is the $\ell$-order moment the GSD of stellar dust from a single generation of stars and $X_{{\rm d},\,j}$ is the ``production matrix'' for dust. It is fair to assume the GSD of the dust proved by a specific kind of stars is invariant over time. The net stellar-dust yield may vary with metallicity and thus change over time, but the average GSD is the result of processes that should not depend strongly on metallicity and other time-dependent factors.  However, the stellar GSD should depend on the mass of the star for several reasons, but most notably because SNe and AGB stars produce different types of dust under quite different conditions. It is likely that dust production in SNe and AGB stars give rise to distinctly different GSDs, but also reasonable to assume these GSDs are the same regardless of the initial mass and metallicity of the SN progenitor or the low- and intermediate-mass star reaching the AGB. The stellar-mass dependence of $K_\ell^\star$ is then reduced to a simple conditional switch between two GSDs, occurring at stellar mass somewhere in the range $8-10M_\odot$. Under the IRA, $K_\ell^\star$ may then be assumed to be a constant with respect to time. In case the dust from SNe and AGB stars do not have similar GSDs, the numerical values of $K_\ell^\star$ could depend on the shape the IMF, as the balance between AGB dust and SN dust is much dependent of the IMF.  

In case one adopts the IRA the injection rate then becomes
\begin{eqnarray}
\label{jstar_ira}
\nonumber
\left({d\mathcal{K}_\ell\over dt} \right)_\star &=& {1\over \varsigma_\star}{K_\ell^\star\over K_3^\star} \int_0^\infty \,p_{{\rm d},\,j}(m_\star)\,\phi(m_\star)\, \dot{\Sigma}_{\star}(t)\,dm_\star\\
 &=&  y_{{\rm d},\,{j}}  {1\over \varsigma_\star}{K_\ell^\star\over K_3^\star}\dot{\Sigma}_{\star}(t) ,
\end{eqnarray}
where $y_{{\rm d},\,j}$ is the effective stellar dust yield for a specific dust species $j$.  If the IMF is non-evolving, so is the stellar GSD as well, which means one does not need to have different GSDs depending on stellar mass. In such case, $y_{{\rm d},\,j}$ and $K_\ell^\star$ are constants with respect to time. In the simplistic toy models to be presented below, it is therefore assumed that stellar dust follows a power-law GSD with the canonical slope derived for interstellar dust by \citet{MRN77}, which is similar to the GSDs obtained in simulations of SN dust production \citep[see, e.g.,][]{Nozawa03}. 

Interstellar dust is also ``swallowed" in the process of forming new stars, which is an important dust-destruction mechanism.
The destruction rate is of course directly defined by the star-formation rate, which means that the effect on $\mathcal{K}_\ell$ is regulated by the star-formation timescale $\tau_{\rm sf}$, i.e.,
\begin{equation}
\left( {d \mathcal{K}_\ell\over dt}\right)_{\rm astr} = {\mathcal{K}_\ell(t)\over \tau_{\rm sf}(t)}, \quad {1\over\tau_{\rm sf}(t)} = {\dot{\Sigma}_\star(t)\over \Sigma_{\rm gas}(t)}.
\end{equation}
The long-term result of this type of dust destruction can be seen in evolved galaxies, such as ellipticals, which have little interstellar gas and dust due to previous star formation.

\subsubsection{Dust condensation in molecular clouds}
The interstellar gas consists of a significant fraction of molecules, which are formed by gas-phase processes as well as in reactions on the surfaces of dust grains.
In molecular clouds (especially the cores) cold grains may grow by condensation (accretion of specific molecules which are the ``growth species'' for the type of dust present) due to high molecular densities. Moreover, because of the low grain temperatures and the shielding from radiation due to molecular opacity, the sublimation rate for these grains may be regarded as negligible. Thus, effectively, the ``advection term'' of Eq. (\ref{eoc}) gives rise to a pure growth term in Eq. (\ref{momeq}). It should be stressed that the dust condensation model presented below does not describe the local rate of condensation in the high density regions (clouds) the cold ISM phase, but rather the overall rate of condensation in the ISM as a whole, similar to in the models by \citet{Calura08,Dwek98,Pipino11,Valiante11}.

The mass-accretion rate onto a spherical grain of radius $a$ in the free-molecular regime is given by 
\begin{equation}
{dm\over dt} = 4\pi\,a^2 \, \alpha_{\rm s}\,\langle v_{\rm mol}\rangle \,\rho_{\rm mol}(t),
\end{equation}
where $\alpha_{\rm s}$ is the sticking probability, i.e., the fraction of molecules hitting the surface of the grain which also stick to the surface, $\langle v_{\rm mol}\rangle $ is the mean relative velocity of the molecules and $\rho_{\rm mol}$ is the space (volume) density of the relevant molecules in the cloud. Since $dm = 4\pi\,a^2\,\rho_{\rm gr}\,da$, where $\rho_{\rm gr}$ is the bulk density of the grain material, the expression for the condensation-growth velocity by accretion of the key species $k$ (e.g., silicon) can be written
\begin{equation}
\xi_{j,\,k}(t) = {da \over dt} = {\alpha_{\rm s}\,\langle v_{\rm mol}\rangle \over \rho_{\rm gr}} \left[{A_{{\rm eff},\,j}\over A_k} {\rho_k(t)} - {\rho_{{\rm d}, j}(t)} \right],
\end{equation}
where $\rho_k$ is the density of the key species $k$, $\rho_{{\rm d}, j}$ is the density of the dust species $j$ and $A_k$ is the atomic weight of $k$ and $A_{{\rm eff},\,j}$ is the effective atomic weight of the grain material of the dust species $j$. This growth velocity is independent of the sizes of the grains, but strongly dependent on the level of dust depletion,
\begin{equation}
\mathcal{D}_{k,\,j}(t) \equiv {A_k \over A_{{\rm eff},\,j}} {\Sigma_{{\rm d},\,j}(t)\over \Sigma_{k}(t)}. 
\end{equation}
Thus,
\begin{equation}
\left({d \mathcal{K}_\ell\over dt}\right)_{\rm cond} =  \ell\,\alpha_{\rm s}\,\langle v_{\rm mol}\rangle \,{\rho_k(t)\over \rho_{{\rm gr},\,k}} \left[\mathcal{D}_{k,\,j}^\infty-\mathcal{D}_{k,\,j}(t) \right]\,{\mathcal{K}_{\ell-1}(t)},
\end{equation}
where $\rho_{{\rm gr},\,k} = \rho_{\rm gr}\,A_k/A_{{\rm eff},\,j}$ is the density of element $k$ in the grain material and $\mathcal{D}_{k,\,j}^\infty$ is the maximally possible depletion, which is here assumed to be 90\%. The rate of increase of the dust-mass density due to condensation becomes 
\begin{eqnarray}
\label{condrate}
\nonumber
\left({d \Sigma_{{\rm d},\,j}\over dt}\right)_{\rm cond} &=&  {\Sigma_{{\rm d},\,j}(t)\over \mathcal{K}_3(t)} \left({d \mathcal{K}_3\over dt}\right)_{\rm cond} = {\Sigma_{{\rm d},\,j}(t)\over \tau_{{\rm cond},\,j}(t)}\\
&= &3\xi_{j,\,k}^\infty {\rho_k(t)\over \rho_{{\rm gr},\,k}} \left[1-{\mathcal{D}_{k,\,j}(t)\over \mathcal{D}_{k,\,j}^\infty} \right]\,{\mathcal{K}_2(t)\over \mathcal{K}_3(t)},
\end{eqnarray}
where $\xi_{j,\,k}^\infty = \eta\,\alpha_{\rm s}\,\langle v_{\rm mol}\rangle\, \mathcal{D}_{k,\,j}^\infty$ is a constant which sets the growth-velocity scale. The model can effectively be paramaterised in terms of a single timescale parameter evaluated at a given reference point in time $t_{\rm ref}$ (assumed to be the onset of disc formation in the example models described in Sect. \ref{exmod}). That is,
\begin{equation}
\label{condtscale}
{1\over \tau_{{\rm cond},\,j}^{\rm ref}} = \xi_{j,\,k}^\infty\, a_{32}^{-1} {\rho_k(t_{\rm ref})\over \rho_{{\rm gr},\,k}}, \quad a_{32} \equiv {\mathcal{K}_3(t_{\rm ref}) \over \mathcal{K}_2(t_{\rm ref})},
\end{equation} 
where all other quantities are defined as previously described. 

The fraction of cold molecular gas $\eta$ is here treated as a constant, but the different phases of the ISM are of course in reality evolving over long timescales. Moreover, molecular clouds (MCs) have a certain lifetime, after which they can be assumed to quickly disperse, which plays an important role in the cycling between different ISM phases. \citet{Zhukovska08} used a model with Poisson distributed lifetimes, while for example \citet{Hirashita11} just assumed a characteristic lifetime, but in practice it does not matter that much how one choses to take the MC lifetimes into account in a one-phase model of the ISM. If the condensation of dust is efficient, which is usually the case inside MCs, then the condensation timescale within a MC is shorter than its characteristic lifetime. In such case the model by \citet{Zhukovska08} is effectively very similar to that introduced above \citep[as well as in][]{Calura08,Dwek98,Pipino11}. In the model above, the timescale $\tau_{\rm cond}$ is roughly of the same order as the ``effective exchange timescale'' introduced by \citet{Zhukovska08}. However, to properly include the effects of cycling between ISM phases, a multi-phase model is needed. Such a model is under development and will be included in future work, but for the overall purpose of the present paper it is not that important.

\subsubsection{Coagulation}
In the cold and warm (neutral) phases of the ISM it is expected that grains can clump together and coagulate. The cores of MCs is a likely place for this process to occur \citep{Hirashita14} because of the high densities, but the right conditions can be found also in the warm ($\sim 100$~K) phase of the ISM. Just as for condensation, a one-phase model of the ISM is not sufficient to fully describe the injection of large grains (aggregates) formed by coagulation (a.k.a. coalescence/aggregation). Only a fraction of the dust is found in environments where coagulation can be efficient and the cycling between different phases is of course not instantaneous. The coagulation problem will in the following be treated in the same way as the condensation problem: by calibration of the parameters of the model such that the cycling timescale is effectively taken into account. 

As opposed to condensation, coagulation of grains is a process under which the total dust mass is conserved, while the number of grains decrease. \citet{Smoluchowski16} derived the original desecrate version of the coagulation equation, which is now commonly known as the Smoluchowski coagulation equation (SCE). The SCE is for natural reasons usually formulated in terms of the grain-mass distribution and takes the form
\begin{eqnarray}
\label{descreteSCE}
\nonumber
\frac{\partial f}{\partial t}&=&\frac{1}{2}\sum^{i-1}_{j=1}
C(m_i-m_j,m_j)\,f(m_i-m_j,t)\,f(m_j,t) - \\
&&\sum^\infty_{j=1}C(m_i,m_j)\,f(m_i,t)\,f(m_j,t),
\end{eqnarray}
where $C$ is the coagulation kernel and $f(m,t)$ is the GMD. Each subscript $i$ (or $j$) refers to a fixed ``mass bin'', i.e., particles of a specific masses. The continuous version of the SCE can be obtained by simply replacing the sums with integrals,
\begin{eqnarray}
\label{contSCE}
\nonumber
{\partial f\over \partial t} &=& {1\over 2}\int_0^{m} C(m-m^{\prime},m^{\prime}) f(m-m^{\prime},t) f(m^{\prime},t)\,dm^{\prime} -  \\
&& f(m,t)\int_0^\infty C(m,m^{\prime})\,f(m^{\prime},t)\,dm^{\prime},
\end{eqnarray}
which is an equation mathematically categorised as a convolution-type integro-differential equation. Assuming stationarity, i.e., $\partial f/\partial t=0$, it is essentially a Volterra equation of the first kind and as such it has a general class of solutions where $C$ can be expressed in terms of $f$ and vice versa \citep{Dubovskii92}.  Without simplifying assumptions, the SCE is an analytically tractable problem only for three specific cases known as the constant, linear and multiplicative solutions, referring to the mathematical form of the corresponding kernels. Thus, application of the SCE usually requires numerical methods. Numerical solution of the SCE is tricky and the best way to avoid discretisation errors (``binning problems'') is to use the method of moments, which reduces the problem to solving a simple system of ordinary differential equations (ODEs).

Using the definition of moments (\ref{Mmom}) one obtains equations of the general form
\begin{equation}
\left({d\mathcal{M}_n\over dt}\right)_{\rm coag} = \int_0^\infty m^n{\partial f\over \partial t}\,dm = \mathcal{I}_\uparrow - \mathcal{I}_\downarrow,
\end{equation}
where the ``source integral'' is
\begin{eqnarray}
\label{me_sink}
\nonumber
 \mathcal{I}_\uparrow &=& {1\over 2}\int_0^\infty m^n \int_0^m C(m-m^{\prime},m^{\prime})\times \\
&& f(m-m^{\prime},t) f(m^{\prime},t)\,dm^{\prime}dm,
\end{eqnarray}
and the ``sink integral'' is
\begin{equation}
\mathcal{I}_\downarrow = \int_0^\infty m^n\,f(m,t)\int_0^\infty C(m,m^{\prime})\,f(m^{\prime},t)\,dm^{\prime}dm.
\end{equation}
By reversal of the integration order in eq. (\ref{me_sink}) and substitution of $\mu = m - m^{\prime}$ and then letting $\mu \to m$, one arrives at 
\begin{eqnarray}
\label{SCEmom}
\nonumber
\left({d\mathcal{M}_n\over dt}\right)_{\rm coag} &=& {1\over 2} \int_0^\infty \int_{0}^\infty C(m,m^{\prime})\, f(m,t) f(m^{\prime},t)\times \\
&& [(m+m^{\prime})^n-m^n-(m^{\prime})^n]\,dm^{\prime}dm.
\end{eqnarray}
Because $m \propto a^3$, this equation can be rewritten in terms of the grain radii $a$, $a^{\prime}$ and thus provide equations for the moments $\mathcal{K}_\ell$,
\begin{eqnarray}
\label{SCEmom2}
\nonumber
\left({d\mathcal{K}_\ell\over dt}\right)_{\rm coag} &=& {1\over 2} \int_0^\infty \int_{0}^\infty {C}(a,a^{\prime})\, f(a,t) f(a^{\prime},t)\times \\
&& \left\{[a^3+(a^{\prime})^3]^{\ell/3}-a^{\ell}-(a^{\prime})^{\ell}\right\}\,da^{\prime}da.
\end{eqnarray}
A characteristic timescale for coagulation can be defined in terms of the change in the number density of grains $n_{\rm d}$ (the zeroth moment $\mathcal{K}_0$),
\begin{equation}
\label{tau_coag}
{1\over \tau_{\rm coag}(t)} = {1\over n_{\rm d}}\left|{dn_{\rm d} \over dt}\right|_{\rm coag}. 
\end{equation}
A reference timescale analogous to that discussed above for condensation can be defined by evaluating $n_{\rm d}$ at $t=t_{\rm ref}$.

Kernels ${C}$ which are expressible as a combination of powers of the grain radii $a$ and $a^{\prime}$, will generate exact equations for moments of order $\ell = 0,3,6,\dots$, since the right hand side will then only contain moments of $f(a,t)$. For other orders ($\ell = 1,2,4,5,\dots$), an approximation to obtain closure is necessary. Moreover, that the kernel allows exact equations is not always the case, which then adds another level of difficulty in obtaining closure of the moment hierarchy. The simplest (also the most stable) method to close the moment hierarchy is to apply a well-calibrated interpolation scheme to approximate the ``intermediate moments'' while solving the system of moment equation for $\ell = 0,3,6,\dots$ numerically. This method is discussed in more detail in Section \ref{numerical}.

\subsubsection{Fragmentation}
\label{fragmentation}
If the relative velocity between two colliding dust grains is high enough, the result may be fragmentation of one, or both, of the particles. There exist a transitional collision-energy range, which marks the regime where the result of grain--grain interactions is no longer just coagulation or scattering \citep{Jones96}. This regime is more easily accessed if the gaseous medium, in which the dust grains reside, is intrinsically turbulent or is hit by a shock wave (which may also induce turbulence). Fragmentation can thus be associated with SN events \citep{Jones96}, but in general it is not clear to what extent fragmentation occur in the ISM. The fragmentation model described below is included in the code, but is not used in the toy models presented in Section \ref{exmod}, since main purpose of those models is to demonstrate the effects of coagulation on the condensation rate. 

In analogy with the SCE, the fragmentation equation can be written \citep{Dubovskii92}
\begin{eqnarray}
\label{FEQ}
\nonumber
{\partial f\over \partial t} &=& -{f(m,t)\over m}\int_0^m m^{\prime}\gamma(m,m^{\prime})\,dm^{\prime} +\\
&&  \int_m^\infty \gamma(m^{\prime},m)\,f(m^{\prime},t)\,g(m^{\prime})\,dm^{\prime},
\end{eqnarray}
where $\gamma$ is the fragmentation kernel (mass flux through each ``mass bin'') and $g(m)$ is the debris-mass distribution. The first term describes the removal of grains of mass $m$ due to break-up following destructive grain--grain interactions and the second term is a source term for the production of debris. Following \citet{Vigil89}, the first term can also be written
\begin{equation}
{f(m,t)\over m}\int_0^m m^{\prime}\gamma(m,m^{\prime})\,dm^{\prime} = A(m)\,f(m,t),
\end{equation}
where $A(m)$ is the break-up rate of particles with mass $m$. The second term can be approximated
\begin{eqnarray}
\nonumber
\int_m^\infty \gamma(m,m^{\prime})\,f(m^{\prime},t)\,g(m^\prime)\,dm^{\prime}  \approx &\\
\int_m^\infty A(m^{\prime})\,B(m^{\prime}|\, m)\,f(m^{\prime},t)\,dm^{\prime}
\end{eqnarray}
where $B(m\, |\, m^{\prime})$ is a number distribution function describing the probability of producing a particle of mass $m$ from the break-up of a particle of mass $m^{\prime}$ (a.k.a. the `daughter rate'). The corresponding (approximate) fragmentation equation is then
\begin{equation}
{\partial f\over \partial t} \approx -A(m)\,f(m,t) + \int_m^\infty A(m^{\prime})\,B(m\, |\, m^{\prime})\,f(m^{\prime},t)\,dm^{\prime}.
\end{equation}
This formulation of the fragmentation problem is in several aspects more convenient than eq. (\ref{FEQ}), e.g., because break-up can be induced in more than one way, it may be difficult to find a representative kernel. The functions $A(m)$ and $B(m\, |\, m^{\prime})$ can parameterised and calibrated to experimental results such that one can write down equations for all moments. That is, 
\begin{eqnarray}
\nonumber
\left({d\mathcal{M}_{n}\over dt}\right)_{\rm frag} &=& \int_0^\infty m^{n}{\partial f\over \partial t}\,dm \\\nonumber
&=& -\int_0^\infty m^{n} A(m)\,f(m,t)\,dm + \\\nonumber
&& \int_0^\infty m^{n}\int_m^\infty A(m^{\prime})\,B(m\, |\, m^{\prime})\,f(m^{\prime},t)\,dm^{\prime}dm \\
&=& -I_n(t) + J_n(t),
\end{eqnarray}
where mass conservation is ensured by adding the auxiliary condition
\begin{equation}
\label{masscons}
m^{\prime} = \int_0^{m^{\prime}} m\,B(m\,|\, m^{\prime})\,dm.
\end{equation}
The distribution function $B(m\, |\, m^{\prime})$ should be a homogenous function and expressed in terms of the ratio $m^{\prime}/m$. That is, $B(m\, |\, m^{\prime}) = b(m/m^{\prime})/m^{\prime}$, which makes it possible to find an expression for the second term $J_n$,
\begin{eqnarray}
\nonumber
J_n (t) 
&=&  \int_0^\infty {A(m^{\prime})\over m^{\prime}} \,f(m^{\prime},t)\,dm^{\prime} \int_0^{m^{\prime}} m^{n} b(m/m^{\prime})\, dm \\\nonumber
&=& \int_0^\infty m^{n} A(m)\,f(m,t)\,dm\int_0^1 z^{n} b(z)\,dz \\
&=& I_n(t)\,L_n,
\end{eqnarray}
where $z \equiv m/m^{\prime}$. The moment of $b(z)$, denoted $L_n$ above, must satisfy mass conservation, i.e., the zeroth moment must equal the average number of fragments produced in a break-up event and the first moment ($n = 1$) must be unity, which follows from equation (\ref{masscons}). The moment equations can thus be written on compact form as
\begin{equation}
\left({d\mathcal{M}_{n}\over dt}\right)_{\rm frag} = (L_n - 1)\,I_n(t).
\end{equation}
For so-called  ``power-law break-up'', with $A(m) = k_0\, m^\mu$, $b(m/m^{\prime}) = (\nu+2)\,(m /m^{\prime})^{\nu}$, one then obtains
\begin{equation}
\left({d\mathcal{M}_n\over dt}\right)_{\rm frag} = k_0 \left({1-n \over n + \nu + 1}\right) \mathcal{M}_{n + \mu}, \quad \nu \le -2,
\end{equation}
which is a commonly used model for linear fragmentation \citep{Brilliantov15,Vigil89}. In terms of the grain-size moments $K_\ell$ an analogous equation is
\begin{equation}
\left({d\mathcal{K}_\ell\over dt}\right)_{\rm frag} = \tilde{k}_0(\mu) \left({1-\ell/3 \over \ell/3 + \nu + 1}\right) \mathcal{K}_{\ell + 3\mu}, \quad \nu \le -2,
\end{equation}
where $\tilde{k}_0(\mu) = (4\pi\,\rho_{\rm gr}/3)^{3\mu}k_0$.

\subsubsection{Dust destruction by sputtering}
The dominant dust-destruction mechanism in the ISM is thought to be sputtering induced by the passage of SN shock waves. The most commonly used model for such dust destruction is the simple prescription according to \citet{McKee89},  which results in a dust-destructions timescale given by
\begin{equation}
{1\over\tau_{\rm d}(t)} = \langle M_{\rm ISM}\rangle\, {R_{\rm SN}(t)\over \Sigma_{\rm gas}(t)},
\end{equation}
where $R_{\rm SN}$ is the SN rate per pc$^2$ and $\langle M_{\rm ISM}\rangle$ is the average mass of interstellar gas affected by each supernova event, a number which is typically $10^3\,M_\odot$ (this number is assumed in all model examples that follows in the present paper). 

As explained in \citet{Asano13b} a generalised dust-destruction timescale must be considered to be dependent on the GSD. Sputtering may affect grains of different sizes in different ways, or at least remove a larger fraction of the mass from a small grain compared to a large grain. Hence, one may generalise the expression for the destruction timescale into
\begin{equation}
{\tau_{\rm d}(t)\over\tau_{\rm d,\ell}(t)} = \left[1- {1\over \mathcal{K}_\ell(t)}  \int_{0}^{\infty}a^{\ell}\int_{0}^{\infty} f(a^{\prime},t)\,\theta(a^{\prime}|\, a)\,da^{\prime}da\right],
\end{equation}
where the function $\theta(a^{\prime}|\, a)$ will be referred to as the "transfer function" and describes how the GSD changes after the passage of a SN shock. The transfer function $\theta$ should be essentially time-independent throughout the course of evolution of the ISM and have only a weak dependence on the GSD. As in the work of \citet{Asano13b} $\theta(a'\to a)$ is assumed to be a simple function of grain radius only. Moreover, it is assumed that the convolution of $\theta(a'\to a)$ and $f(a,t)$ can be separated into a function $\mathcal{H}(a,t)$ and $f(a,t)$, i.e.,
\begin{equation}
\int_{0}^{\infty} f(a',t)\,\theta(a^{\prime}|\,a)\,da' = f(a,t)\,\mathcal{H}(a,t),
\end{equation}
which yields
\begin{equation}
{\tau_{\rm d}(t)\over\tau_{\rm d,\ell}(t)} = \upsilon \left[1- {1\over \mathcal{K}_\ell(t)}  \int_{0}^{\infty}a^{\ell}  f(a,t)\,\mathcal{H}(a,t)\,da\right],
\end{equation}
where $\upsilon$ is an arbitrary scaling parameter used to calibrate the model such that the reduction in dust mass over a Galactic age is similar to that assumed above ($10^3\,M_\odot$) for use with the prescription by \citet{McKee89}. This calibration is {\it ad hoc} and not self-evident, however. Observational constraints (e.g., extinction curves) may suggest different overall levels of dust destruction depending on the shape of the GSD. Note that in the above model of dust-destruction, the timescale is not generally the same for all moments $\mathcal{K}_\ell$. In the simplest case, $\mathcal{H}(a,t) = 0$, i.e., all dust grains in the affected volume of the ISM is completely sputtered away (the other trivial case is $\mathcal{H}(a,t)=1$, which corresponds to no dust destruction due to sputtering). In such case, the timescale is the same for all $\mathcal{K}_\ell$, and the simple formula according to \citet{McKee89} is recovered. Note that $\langle M_{\rm ISM}\rangle$ should be seen as the corresponding mass of interstellar gas {\it effectively} cleared from dust by each supernova event on average, while the local efficiency of destruction may vary due to different physical circumstances. 

There is another special case for which an approximate closed form expression can be obtained: a constant sputtering rate $S$ (per unit area), where the sputtering rate is low enough not to affect the largest grains much \citep[which are expected to remain largely unaffected according to simulations, see, e.g.,][]{Slavin04,Bocchio14}. In such case $da/dt = - S/\rho_{\rm gr}$ for all grain radii $a$. Since the time a grain is subject to SN shock induced sputtering should be approximately the same for all grains ($\approx$ constant shock speed) all grains would decrease in size by an effective amount $\Delta a$. Thus, 
\begin{equation}
{\tau_{\rm d}(t)\over\tau_{\rm d,\ell}(t)} = \upsilon \left[1-  {\mathcal{H}_0\over \mathcal{K}_\ell(t)}  \int_{0}^{\infty}a^{\ell}  f(a-\Delta a,t)\,da\right],
\end{equation}
which corresponds to $\mathcal{H}(a,t) = \mathcal{H}_0\,f(a-\Delta a,t)/f(a,t)$. The largest grains are left almost unaffected, so one can safely assume $a_{\rm max}\gg \Delta a$, so that $a_{\rm max} \pm \Delta a\approx a_{\rm max}$.  If one then let $x = a-\Delta a$, $a_{\rm max} \to \infty$ and require that $f(a,t)=0$ for $a \leq \Delta a$, one may then write 
\begin{equation}
{\tau_{\rm d}(t)\over\tau_{\rm d,\ell}(t)} \approx  \upsilon \left[1- \mathcal{H}_0 \sum_{k=0}^\ell {\ell!\over k!(\ell-k)!} {\Delta a^k \mathcal{K}_{\ell-k}(t)\over \mathcal{K}_\ell(t)} \right].
\end{equation}
This is under most circumstances a sufficiently general model for the sputtering timescale in a model of galactic dust evolution. But\citet{Mattsson14a} suggested that since grain--grain interaction associated with passage of SN shocks may induce fragmentation, the rate of dust destruction by sputtering should depend on the abundance of dust grains. The debris consists of tiny grains which are easily sputtered away completely, while large grains can escape essentially unharmed, as shown by \citet[e.g.,][]{Slavin04}. If this hypothesis is confirmed by direct numerical simulations, it would mean that the destruction timescale is, to first order, inversely proportional to the dust density of the ISM. Consequently, dust destruction due to SN induced sputtering would be less efficient at early times. 

\subsection{A dust-evolution toy model}
\label{exmod}
As an example of a model of interstellar dust processing based on the method of moments, a simple model of the dust evolution in the solar neighbourhood is presented below. These are only first results from the new code and should not be seen as a precise and accurate model of the build-up of the Galactic dust component. However, the method of moments for dust processing in the ISM, as described in previous sections, can quite easily be combined with a detailed chemical evolution model such as that of \citet{Mattsson10b}.

\subsubsection{Adopted equations and calibration of the model}
The equations for the formation and chemical evolution of the Galaxy are those given in Sect. \ref{model}, adopting the IRA and only solving for the total metallicity $Z$ and a single generic dust component rather than for individual elements $i$ and dust species $j$. Adopted parameter values are given in Tables \ref{calibration} and \ref{parameters}. This model is combined with the following equation for the total dust density,
\begin{eqnarray}
\nonumber
{d\Sigma_{\rm d}\over dt} &=& \left[y_{\rm d} - {\Sigma_{\rm d}(t)\over \Sigma_{\rm gas}(t)}\right]{d\Sigma_{\rm stars}\over dt} + \\
&&{\Sigma_{\rm d}(t) \over \mathcal{K}_3(t)}\left[\left({d\mathcal{K}_3\over dt}\right)_{\rm cond} - \left({d\mathcal{K}_3\over dt}\right)_{\rm dest}\right]
\end{eqnarray}
where the ``dust processing terms'' are given in terms of the corresponding rate of change for the moment $\mathcal{K}_3$ (which is proportional to the volumetric dust-mass density). Note that fragmentation is not included in this example model, since it may hide the effect of coagulation on the evolution of the average grain size as well as the condensation rate, which is of primary interest here. To obtain $\mathcal{K}_3$ one has to solve the moment hierarchy, i.e., solve the set of ODEs described in Section \ref{moms}, which requires that the coagulation kernel is specified. The kernel used here is of the form \citep[free-particle regime, assuming a universal thermal speed, see, e.g.,][]{Seinfeld98}
\begin{equation}
C(a,a^{\prime}) = {C_0 \over a_0^2}\,(a + a^{\prime})^2, 
\end{equation}
where $C_0/a_0^2 = \pi\, \tilde{\eta}\,\alpha_{\rm s}^{\rm coag}\,\Delta \bar{v}_{\rm d}$ and $a_0$ is a reference grain size, e.g., the initial mean grain radius. In this definition of $C_0$, $ \alpha_{\rm s}^{\rm coag}$ is the effective (or average) sticking probability in grain--grain interactions, $\tilde{\eta}$ is the fraction of dust undergoing coagulation and $\Delta \bar{v}_{\rm d}$ is the typical velocity difference between the interacting particles, i.e., some weighted average of $\Delta v_{\rm d}(a,a^{\prime}) = |v_{\rm d}(a)-v_{\rm d}(a^{\prime})|$. $C_0$ will here be considered a universal constant, although $\Delta {v}_{\rm d}$ is obviously a function of grain size. By eq. (\ref{SCEmom2}), the coagulation term for the zeroth order moment then becomes
\begin{equation}
\left({d\mathcal{K}_0\over dt}\right)_{\rm coag} = -{C_0\over a_0^2} [\mathcal{K}_2(t) \mathcal{K}_0(t) +  \mathcal{K}_1^2(t) ],
\end{equation}
and (due to the conservation of mass)
\begin{equation}
\left({d\mathcal{K}_3\over dt}\right)_{\rm coag} = 0.
\end{equation}
Orders $\ell = 6, 9, 12, 15$ follow analogously (see Appendix \ref{ODEs} for further details). The remaining (intermediate) orders have no corresponding exact equations when coagulation is included according to eq. (\ref{SCEmom2}), which requires use of an unorthodox numerical technique (see below).

The galaxy formation/evolution model contains several parameters that need to be set by a procedure in which the model is calibrated against observational constraints (see Table \ref{calibration}). For the solar neighbourhood, the present-day density of stars, gas and metals in The Galaxy is known with fairly good precision \citep[see, e.g.][]{Asplund09,Dame93,Dickey93,Holmberg04,McKee15}. The star-formation rate is known to at least the right order of magnitude \citep{Rana91} and the rate of infall is known within a factor of a few. \citep{Braun04}. With the figures given in Table \ref{calibration}, the parameters of the galaxy formation model are severely constrained and one may therefore lock the infall timescale, star-formation timescales and IMF/return fraction to the values given in Table \ref{calibration}. 

The parameters of the dust processing model are not arbitrary either. First, it is well established that the Galactic dust-to-metals (mass) ratio is $\approx 0.5$ and the local number density of dust grains is $\sim 10^{-12}$~cm$^{-3}$ \citep[see][and references therein]{Inoue03,Draine07}. Second, the typical/average grain size in the ISM must be of the order $0.1\,\mu$m \citep{Weingartner01}. Bringing these numbers together, it therefore seems likely that stars produce relatively small grains on average (or at least the grains that reach the ISM are small) and that the grain size of interstellar grains is mainly due to growth by condensation and coagulation. The growth-velocity constant $\xi_0$ can be constrained  for a given stellar dust yield (with a specified stellar GSD), generic dust-destruction timescale/efficiency (which is set by $\langle M_{\rm ISM}\rangle$) and a given material bulk density for the dust. One finds that calibration of $\xi_0$, assuming $\rho_{\rm gr} = 3.0$~g~cm$^{-3}$, translates into a growth velocity of the order $\xi_0 \sim 10^{-15}$~cm~yr$^{-1}$. The constants $\rho_{\rm gr}$ and $\xi_0$ corresponds to a reference timescale at the onset of disc formation $\tau_{{\rm cond}}^{\rm ref}$ (see eq. \ref{condtscale}) which is given in Table \ref{parameters} for each considered model.

The coagulation kernel sets the efficiency of the coagulation by adjustment of the constant $C_0$, which cannot be explicitly constrained from observations. However, in order for coagulation to play a significant role in the interstellar processing of dust, the present-day coagulation timescale $\tau_{\rm coag}(t_0)$ (see eq. \ref{tau_coag}) should be comparable to the condensation timescale $\tau_{\rm cond}(t_0)$.  In terms of physics, changing $C_0$ corresponds to changing the typical relative velocity between grains and/or the fraction of interstellar gas that is found in environments where coagulation is an important growth factor for the dust grains. The dynamics of the ISM also plays an important role \citep[see, e.g.,][]{Hirashita09}, which leaves room for variation of $C_0$. \citet{Hirashita09} used the results of \citet{Yan04} and a threshold velocity $v_{\rm coag}$ (above which particles are unlikely to stick) to prescribe the relative motions of grains in different ISM phases. Because of the threshold at $v_{\rm coag}$, coagulation is expected to be efficient only in molecular clouds (in particular the so-called ``dark clouds'' which are dust rich and very opaque). According to the MHD simulations by \citet{Yan04}, the typical value of $\Delta \bar{v}_{\rm d}$ in molecular clouds is of the order $\Delta \bar{v}_{\rm d} \sim 1$~km~s$^{-1}$, which suggest that $C_0/a_0^2$ is at most a few km~s$^{-1}$. The adopted coagulation efficiencies are given inTable \ref{parameters}, where the reference timescale $\tau_{\rm coag,\,0}$ (at $\tau_{\rm G} = 1.0$~Gyr) is given instead of $C_0$ and $a_0$.

Dust destruction by SN induced sputtering is hard to quantify \citep[see, e.g.,][]{Jones11}. The most commonly adopted calibration \citep[based on the prescription by][]{McKee89} for the solar neighbourhood suggests that $\langle M_{\rm ISM}\rangle = 800\,M_\odot$ (corresponding to a present-day timescale $\tau_{\rm d,\,0} = 0.8$~Gyr) for carbonaceous dust and $\langle M_{\rm ISM}\rangle = 1200\,M_\odot$ ($\tau_{\rm d,\,0} = 0.6$~Gyr) for silicates \citep{Jones94,Tielens94}. Thus, $\tau_{\rm d,\,0} = 0.7$~Gyr is in the present work adopted as a representative figure for the present-day timescale of SN induced dust destruction. To obtain the same level of total dust destruction obtained in one Galactic age when using the simple prescription by \citet{McKee89}, the dimensionless scaling constants $\upsilon$ and $\mathcal{H}_0$ (here, the latter is fixed to unity -- see Table \ref{parameters}), introduced in the prescription assuming a constant sputtering rate (with $\Delta a = 0.001\,\mu$m), must be chosen such that the total amount of dust destruction over one Galactic age $\tau_{\rm G} = 13.5$~Gyr is similar to that obtained from prescription by \citet{McKee89} with $\langle M_{\rm ISM}\rangle = 1000\,M_\odot$. This is based on the assumption that the ``standard calibration'' for the solar neighbourhood is essentially correct, but there is reason to doubt this assumption. First, the assessment by \citet{Jones11} shows how difficult it is to constrain the destruction timescale. Second, as hypothesised by \citet{Mattsson14a}, grain--grain interactions inside the shocks may play an important role as these can lead to shattering/fragmentation of grains and it is well known that small grains are more susceptible to destruction by sputtering \citep[see, e.g.,][]{Slavin04,Bocchio14}.

\subsubsection{Numerical solution}
\label{numerical}
As described above, the method of moments leads to a system of coupled ODEs. Because the coagulation term can only be expressed without approximations for $\ell = 0,3,6,9,\dots$, its is required that one either uses analytical approximation or an unorthodox numerical procedure called interpolative closure. The latter does, in fact, turn out to be the most accurate and stable approach. Closure schemes based on analytical approximations will therefore not be considered here \citep[but see][for one such example]{McGraw97}.

\begin{figure}
\resizebox{\hsize}{!}{\includegraphics[clip=true]{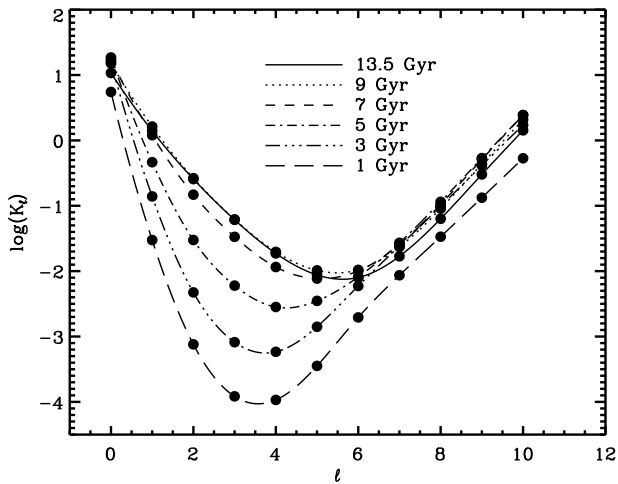}}
\caption{Time evolution of the moment hierarchy up to $\ell = 10$. The black dots show the direct numerical solution of the system of ODEs for the moment hierarchy of model A. Lines show cubic spline fits to the moments of order $\ell = 0, 3, 6, 9$. 
\label{moments}}
\end{figure}

The equations for orders $\ell = 0,3,6,9,\dots$ can be integrated using, e.g., a standard fourth-order Runge-Kutta routine, but since moments of intermediate orders appear on the right-hand side of these equations, a good estimate of the moments of order $\ell = 1,2,4,5,7,8,\dots$ must be obtained at each new time step. Fortunately, the moments $\mathcal{K}_\ell$ show a very predictable trend with moment order $\ell$. Fig. \ref{moments} shows the evolution of the logarithm of the moments (normalised such that $\mathcal{K}_3$ is the dust-mass density) for model A, solving the system of ODEs for {\it all} orders $\ell$ with a fourth-order Runge-Kutta routine up to $\ell = 15$ (tests have shown that the hierarchy can safely be truncated at $\ell = 15$). Cubic spline fits (see below) are shown as solid lines. Clearly, interpolation between the moments of order $\ell = 0,3,6,9,\dots$ can provide accurate estimates of moments of order $\ell = 1,2,4,5,7,8,\dots$, which then provides closure of the system ODEs, albeit with an error that will of course propagate as the numerical solution is advanced from one time step to the next. However, the relative errors are small ($\sim 10^{-4}$ or less) and should not be problem here.

\begin{figure*}
\center
\resizebox{0.73\hsize}{!}{\includegraphics[clip=true]{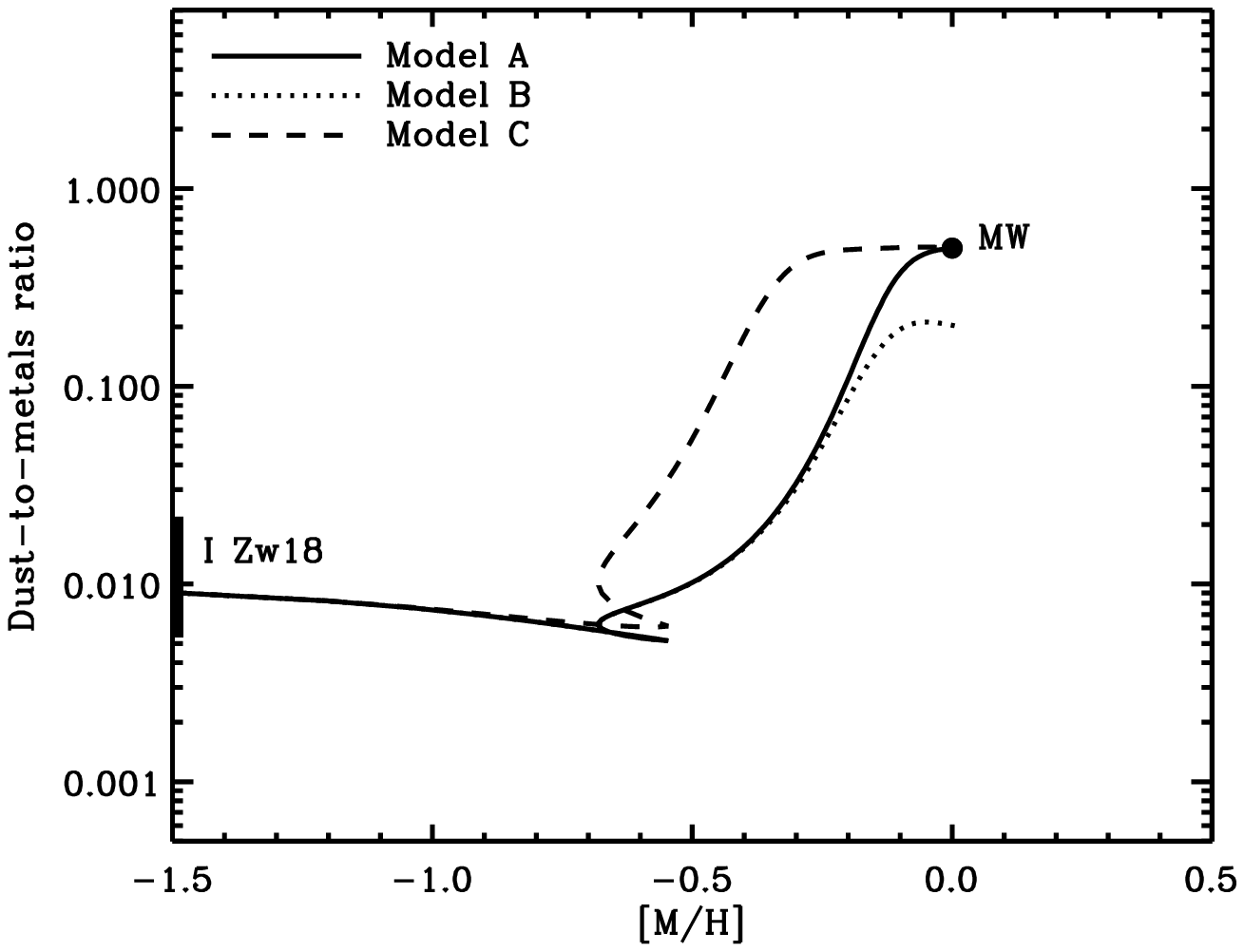}
\includegraphics[clip=true]{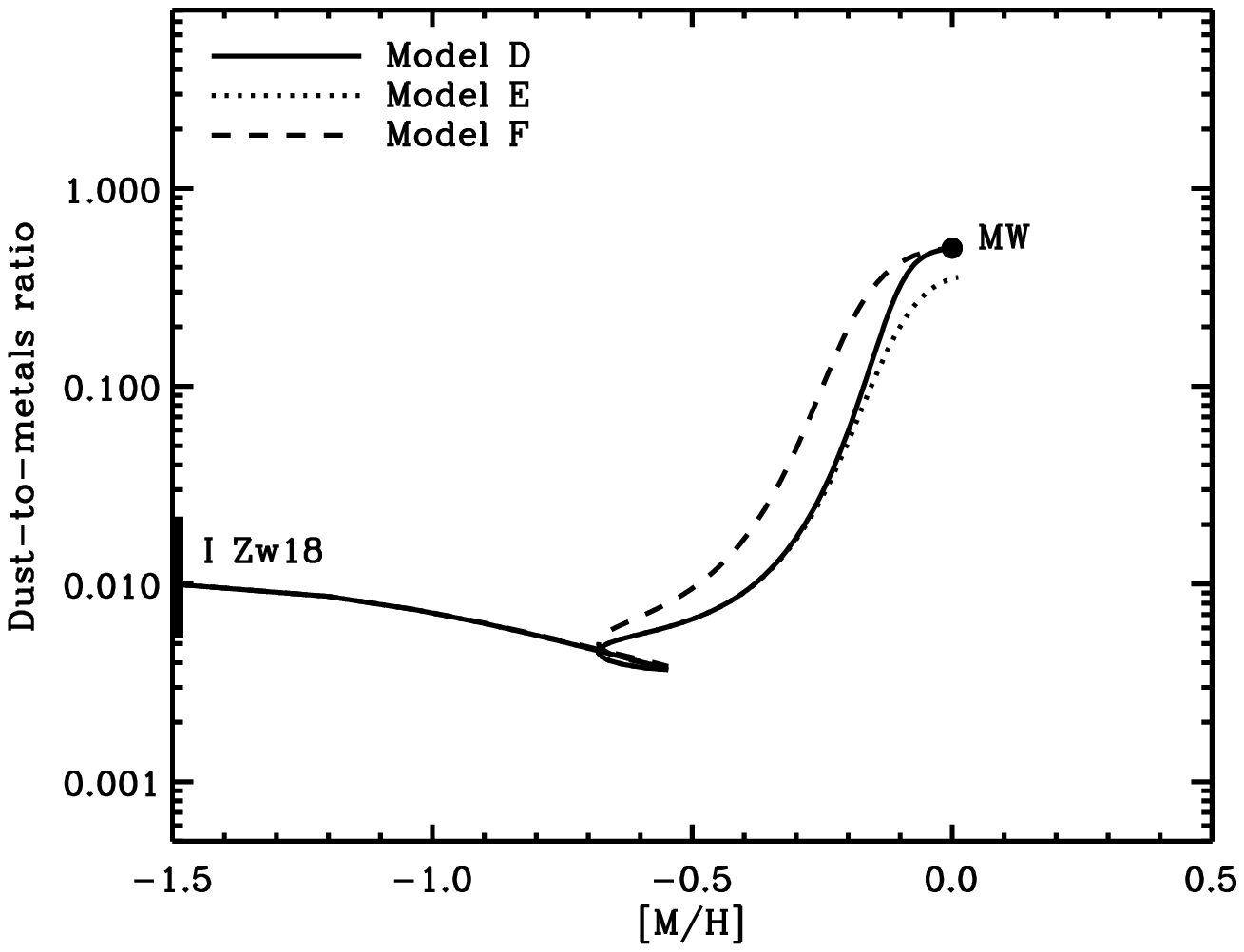}} \\
\resizebox{0.73\hsize}{!}{\includegraphics[clip=true]{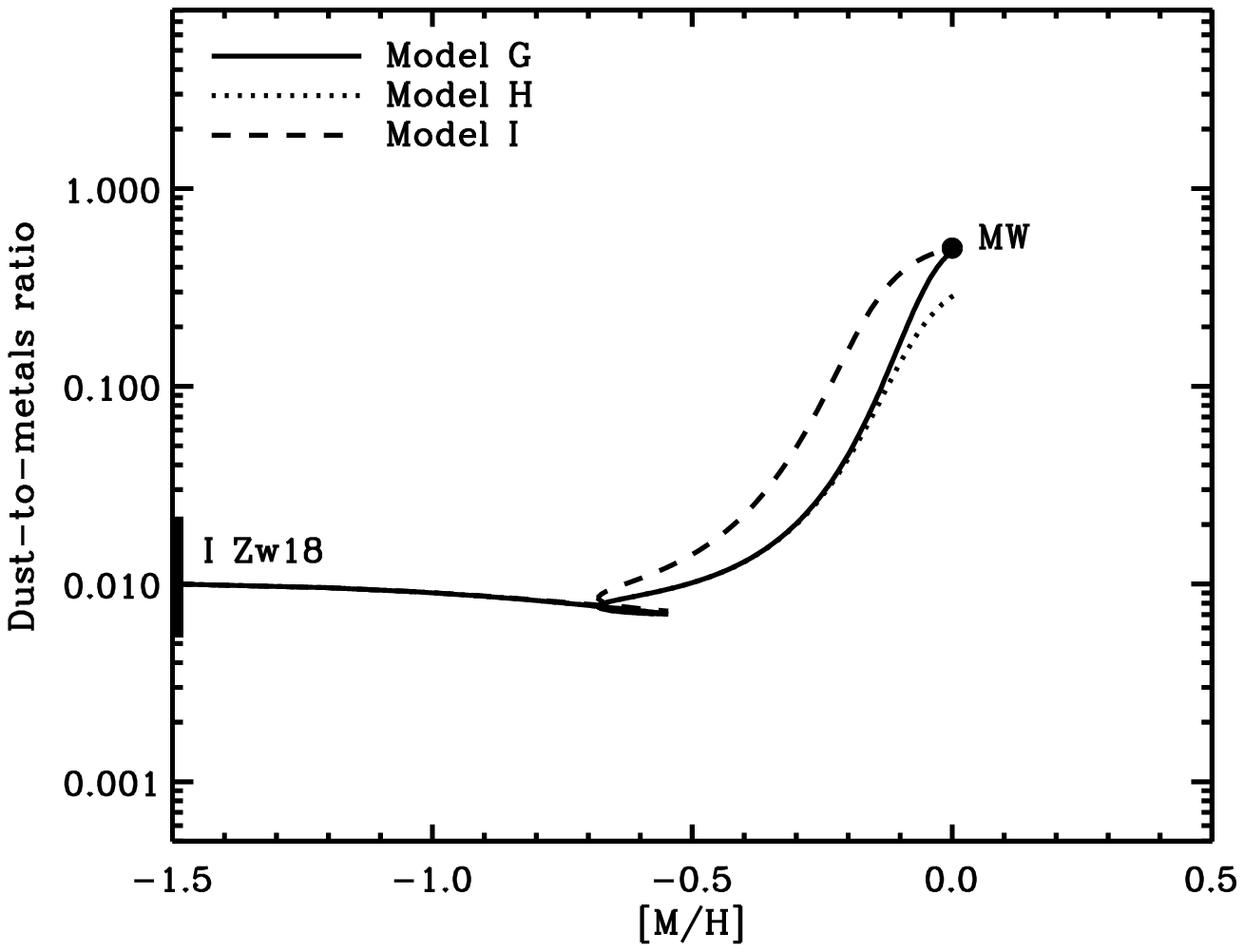}
\includegraphics[clip=true]{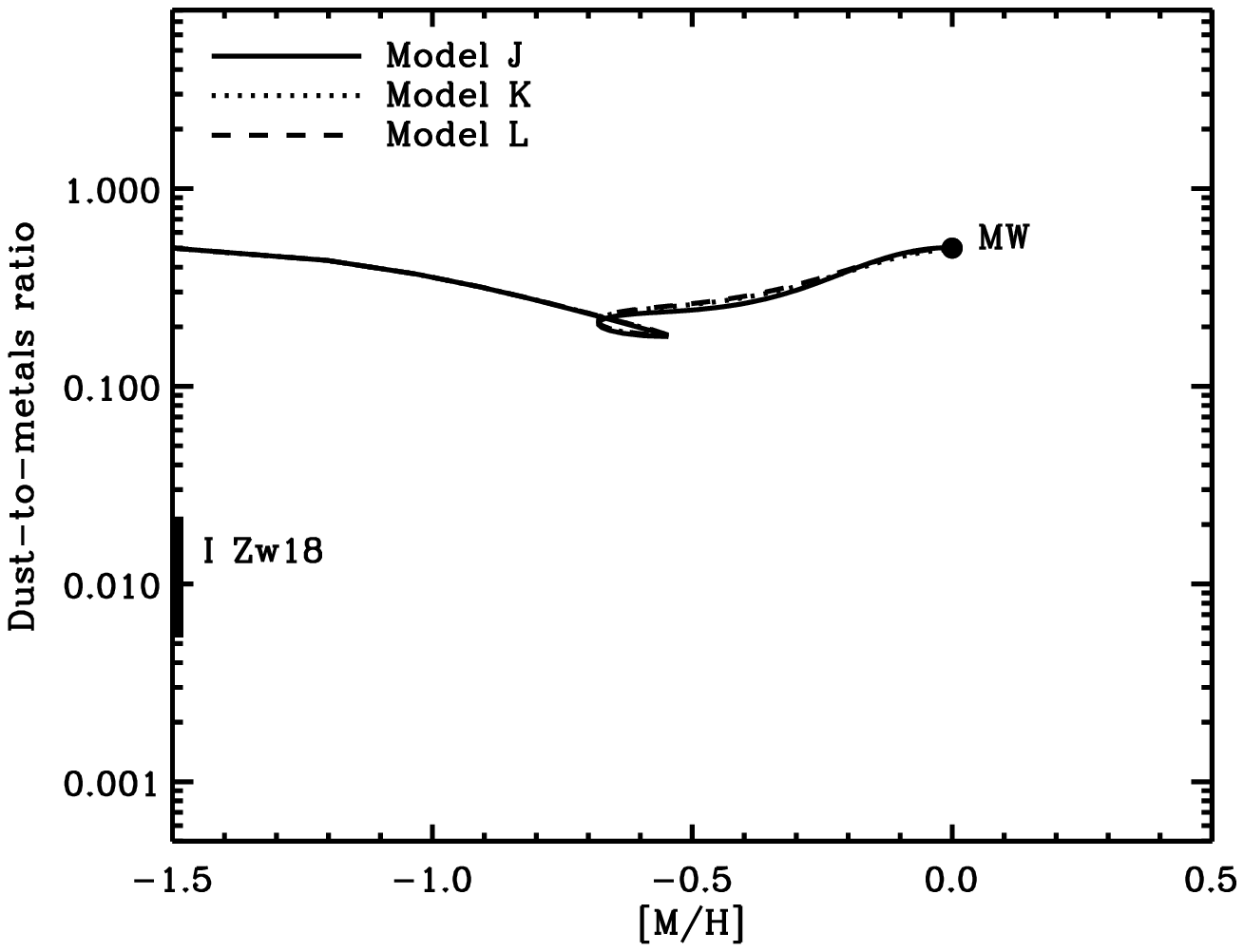}}
\caption{Dust-to-metals ratio as a function of metallicity. Upper left: the simplistic case ($\upsilon = 0$) of dust destruction due to sputtering associated with the passage of SN shock waves. Upper right: the case of a constant sputtering rate and $\Delta a = 0.001\,\mu$m. Lower left: constant sputtering rate with $\Delta a = 0.001\,\mu$m, but for $a^\star_{\rm min} = 0.05\,\mu$m, i.e., a case where stars inject relatively large grains. Lower right: a case with very substantial ($y_{\rm d} = 0.5\,y_Z$) stellar dust production.
\label{MW_sol_test}}
\end{figure*}

Solving the coagulation problem using the method of moments with interpolative closure (MOMIC) is not new \citep[see, e.g.,][]{Frenklach87}, but using the MOMIC for dealing with several types of dust/particle processing and the evolution of the ambient medium at once seems to be almost untouched territory.  \citet{Frenklach87} suggested Lagrange interpolation (a type of polynomial interpolation), which worked reasonably well in the context of their study. In the context of the present paper, however, it turns out to be insufficient in terms of precision and stability. Spline interpolation, on the other hand, has proven to work without causing any numerical problems to speak of. It requires fine tuning of the ``stiffness'' parameter $s$, but extensive testing has shown that $s\to 0$ yields the best result, i.e., a very small value of $s$ is favoured. Small $s$ corresponds to cubic spline interpolation and a cubic spline routine was therefore implemented as the standard method in the code. It is worth noting that a large value of $s$ will result in case which is close to a polynomial interpolation. Thus, it seems the choice of Lagrange interpolation made by \citet{Frenklach87} was not optimal.

  \begin{table*}
  \small
  \begin{center}
  \caption{\label{parameters} Dust-related parameter values of the example models.}
  \begin{tabular}{cccccccccc}
  \hline 
   Model  &  $y_{\rm d}$ & $a_{\rm min}^\star$ & $a_{\rm max}^\star$ & $\Delta a$ & $\mathcal{H}_0$ & $\upsilon$  & $\langle M_{\rm ISM}\rangle$ &  $\tau_{\rm cond}^{\rm ref}$  &  $\tau_{\rm coag}^{\rm ref}$  \\
               &    &  ($\mu$m) & ($\mu$m) & ($\mu$m) &  &  & $M_\odot$ & (Gyr) &  (Gyr)  \\
   \hline
  A  & $1.2\,10^{-4}$ & $5.0\,10^{-3}$ & 5.0 & -              & -   & 0    & $1.0\,10^{3}$ & $1.73$   &     -             \\
  B  & $1.2\,10^{-4}$ & $5.0\,10^{-3}$ & 5.0 & -              & -   & 0    & $1.0\,10^{3}$ & $1.73$   &   $3.48\,10^{2}$  \\
  C  & $1.2\,10^{-4}$ & $5.0\,10^{-3}$ & 5.0 & -              & -   & 0    & $1.0\,10^{3}$ & $0.70$   &   $2.96\,10^{2}$  \\[1mm]
  D  & $1.2\,10^{-4}$ & $5.0\,10^{-3}$ & 5.0 & $1.0\,10^{-3}$ & 1.0 & 83.0 & $1.0\,10^{3}$ & $1.60$   &     -             \\
  E  & $1.2\,10^{-4}$ & $5.0\,10^{-3}$ & 5.0 & $1.0\,10^{-3}$ & 1.0 & 83.0 & $1.0\,10^{3}$ & $1.60$   &   $1.01\,10^{3}$  \\
  F  & $1.2\,10^{-4}$ & $5.0\,10^{-3}$ & 5.0 & $1.0\,10^{-3}$ & 1.0 & 83.0 & $1.0\,10^{3}$ & $1.19$   &   $9.71\,10^{2}$  \\[1mm]
  G  & $1.2\,10^{-4}$ & $5.0\,10^{-2}$ & 5.0 & $1.0\,10^{-3}$ & 1.0 & 83.0 & $1.0\,10^{3}$ & $3.39$   &     -             \\
  H  & $1.2\,10^{-4}$ & $5.0\,10^{-2}$ & 5.0 & $1.0\,10^{-3}$ & 1.0 & 83.0 & $1.0\,10^{3}$ & $3.39$   &   $0.63$          \\
  I  & $1.2\,10^{-4}$ & $5.0\,10^{-2}$ & 5.0 & $1.0\,10^{-3}$ & 1.0 & 83.0 & $1.0\,10^{3}$ & $2.18$   &   $0.61$          \\[1mm]
  J  & $6.0\,10^{-3}$ & $5.0\,10^{-3}$ & 5.0 & $1.0\,10^{-3}$ & 1.0 & 83.0 & $1.0\,10^{3}$ & $2.26$   &     -             \\
  K  & $6.0\,10^{-3}$ & $5.0\,10^{-3}$ & 5.0 & $1.0\,10^{-3}$ & 1.0 & 83.0 & $1.0\,10^{3}$ & $2.36$   &   $8.62\,10^{2}$  \\
  L  & $6.0\,10^{-3}$ & $5.0\,10^{-3}$ & 5.0 & $1.0\,10^{-3}$ & 1.0 & 83.0 & $1.0\,10^{3}$ & $2.30$   &   $8.61\,10^{2}$  \\[1mm]
  \hline
  \end{tabular}
  \end{center}
  \end{table*}

\section{Results}
\label{results}
\subsection{The build-up of the Galactic dust component}
In agreement with previous studies \citep[e.g.,][]{Asano13a,Asano13b, Zhukovska08}, the fraction of metals in dust grains (dust-to-metals ratio) increases promptly at a specific metallicity (see Fig. \ref{MW_sol_test}) and accidentally coincides in time with the onset of disc formation (which is the reason for the "epicycles" in Fig. \ref{MW_sol_test}). The rapid increase is due to the existence of a critical metal density, which may be fairly universal. Strictly speaking, however, there is no universal critical {\it metallicity}, since the metallicity also depends on the gas-mass fraction.  At which metal density this rapid increase occurs, is due to essentially three factors: the sticking probability $\alpha_{\rm s}$, the material density $\rho_{\rm gr}$ and the average relative speed $\langle v_{\rm mol}\rangle$ of the molecules hitting the dust grain. Since in the present toy model it is assumed that $\alpha_{\rm s} = 1$ and $\rho_{\rm gr} = 3.0$~g~cm$^{-3}$, the only remaining parameter is $\langle v_{\rm mol}\rangle$. In models C, F, I and L $\langle v_{\rm mol}\rangle$ is increased by a certain factor, chosen to compensate for the increase of the condensation timescale (which is the reason for the lower present-day values of the dust-to-metals ratio in models B, E, H, and K) when coagulation is introduced (see below for an extended discussion). Using the simple dust-destruction prescription by \citet{McKee89} and a coagulation timescale as given in Table \ref{parameters} (Model B), $\langle v_{\rm mol}\rangle$ must be increased by a factor $\approx 8/3$ compared to the case without  coagulation (compare models A and C, see Fig. \ref{MW_sol_test} and Table \ref{parameters}). As a result of the increased condensation efficiency, the critical metallicity is lowered $\sim 0.2$~dex. 

The early evolution of the dust-to-metals ratio shows a slightly declining trend. This due to the rate of dust destruction by star formation and sputtering induced by shock waves from SNe, which is always exceeding the stellar dust-production rate at some point, unless the stellar dust yield increases with time. The phenomenon has been demonstrated by analytical work \citep[see, e.g.,][]{Mattsson11b} and the declining trend seen in Fig. \ref{MW_sol_test} is thus quite expected. Neither a high stellar dust yield (see Fig. \ref{MW_sol_test}, lower right panel), nor an increased grain size (see Fig. \ref{MW_sol_test}, lower left panel) changes this trend, as expected. The high-yield case shows also that coagulation in the ISM only has a minor effect on the evolution of the dust-to-metals ratio when stars dominate the production of new dust.

\begin{figure*}
\center
\resizebox{0.73\hsize}{!}{\includegraphics[clip=true]{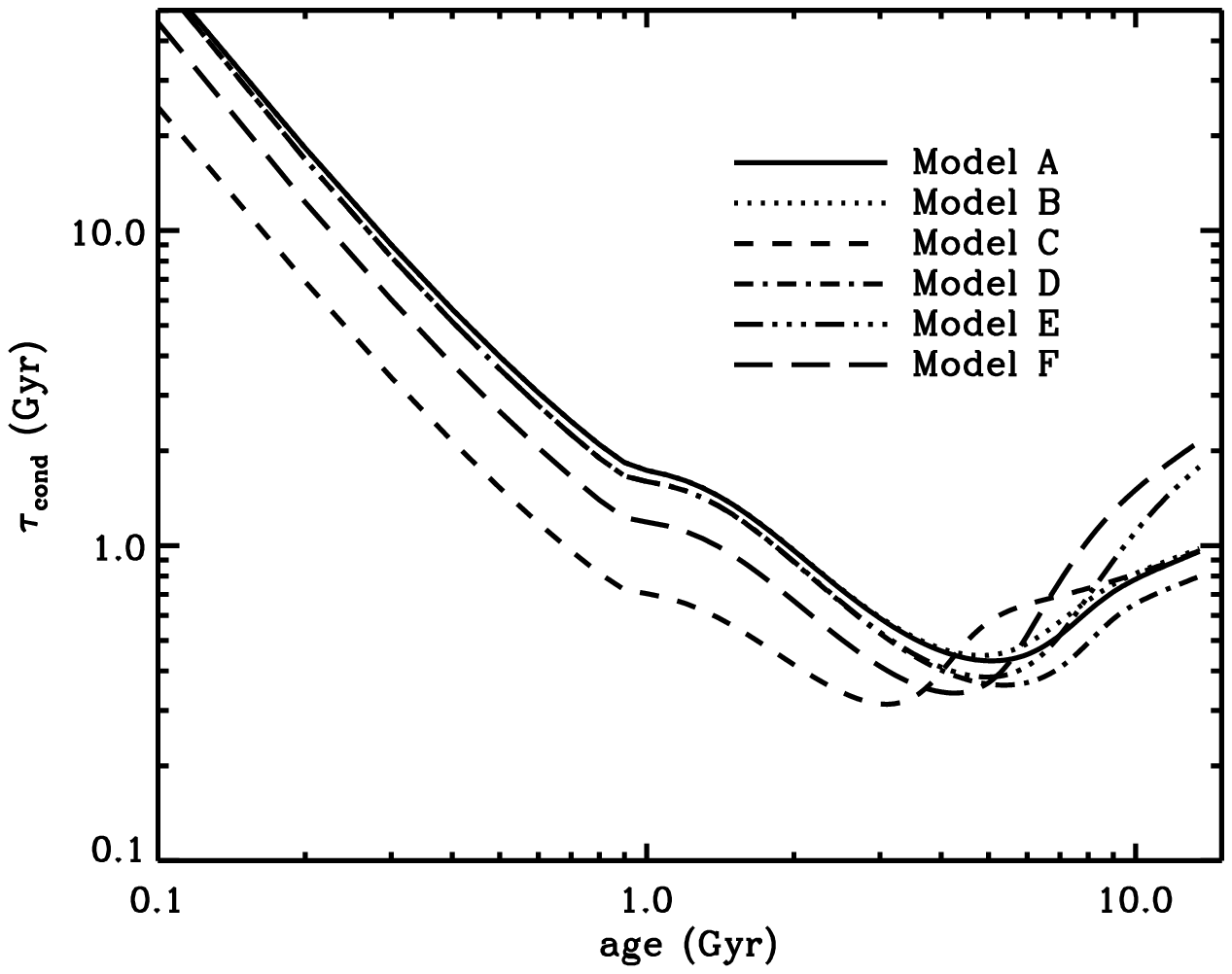}
\includegraphics[clip=true]{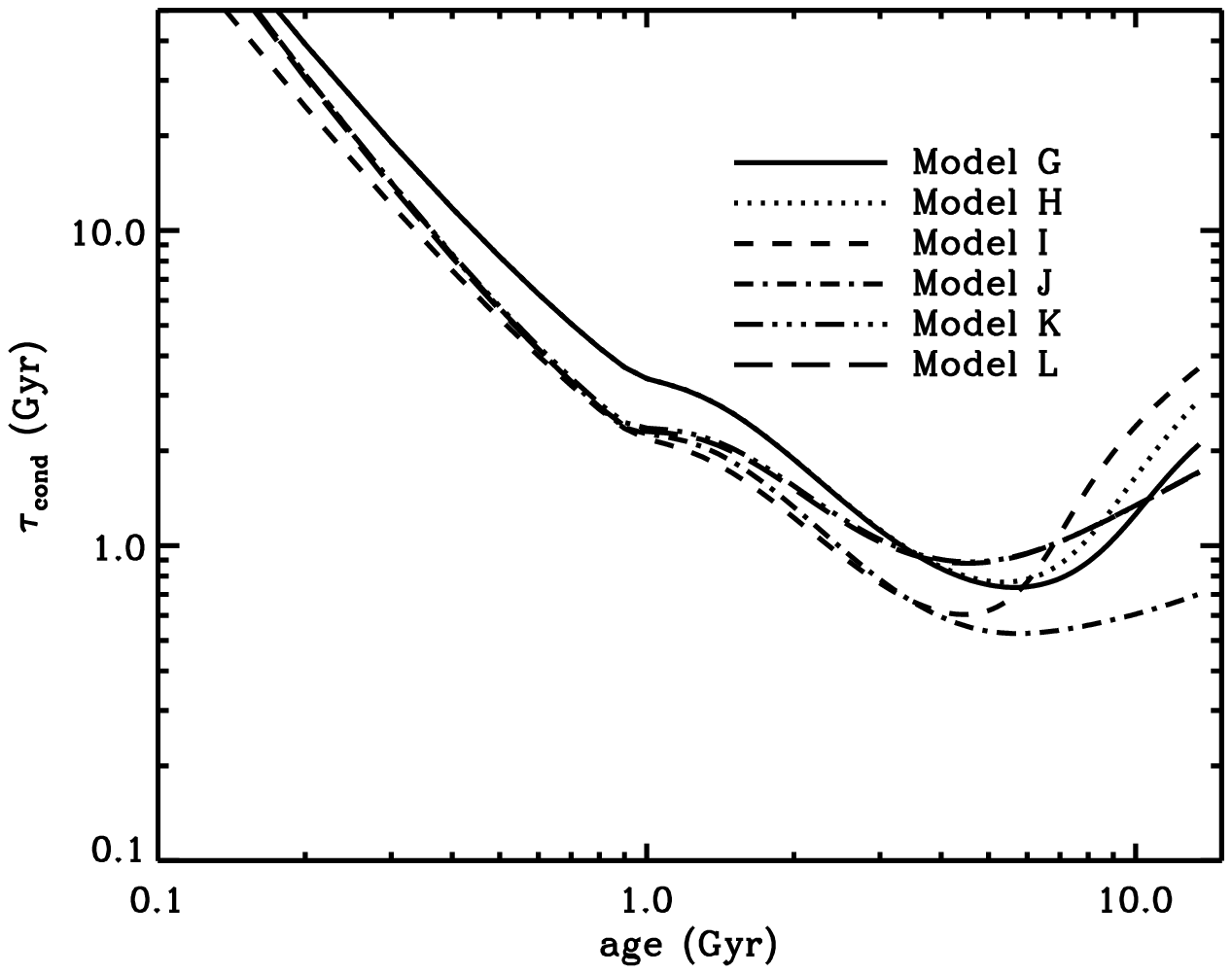}}
\caption{Evolution of the dust-condensation timescale. The left panel shows models A--F and the right panel models G--L.
\label{MW_sol_tau}}
\end{figure*}

\subsection{Evolution of the dust-condensation timescale}
The differences in condensation efficiency at late stages between cases with and without coagulation (e.g., models A and B vs. models D and E), as well as the two considered different prescriptions for dust destruction by sputtering, are largely due to the differences in the volume/area-relation. In a dust population dominated by very small grains, the total surface area of the grains is significantly larger than for a population of normal-sized grains of a the same total mass. Since the condensation rate is proportional to the second moment $\mathcal{K}_2$ (which is proportional to the total surface area of the grains), changes in the average grain size will affect the condensation timescale. Clearly, adding coagulation to the picture can have a non-negligible effect on the condensation timescale (although the overall net efficiency in these example models may be somewhat extreme).  The effect is also amplified by the introduction of a grain-size dependent sputtering prescription (see Fig. \ref{MW_sol_tau}).

It is worth mentioning, once again, that none of the models take interstellar fragmentation into account, although the code is capable of handling fragmentation within the framework presented in Sect. \ref{fragmentation}. The inclusion of fragmentation could counteract the influence that coagulation has on the average grain size and the condensation rate. 

\subsection{Number density and grain-size distribution (GSD)}
\begin{figure*}
\center
\resizebox{0.73\hsize}{!}{\includegraphics[clip=true]{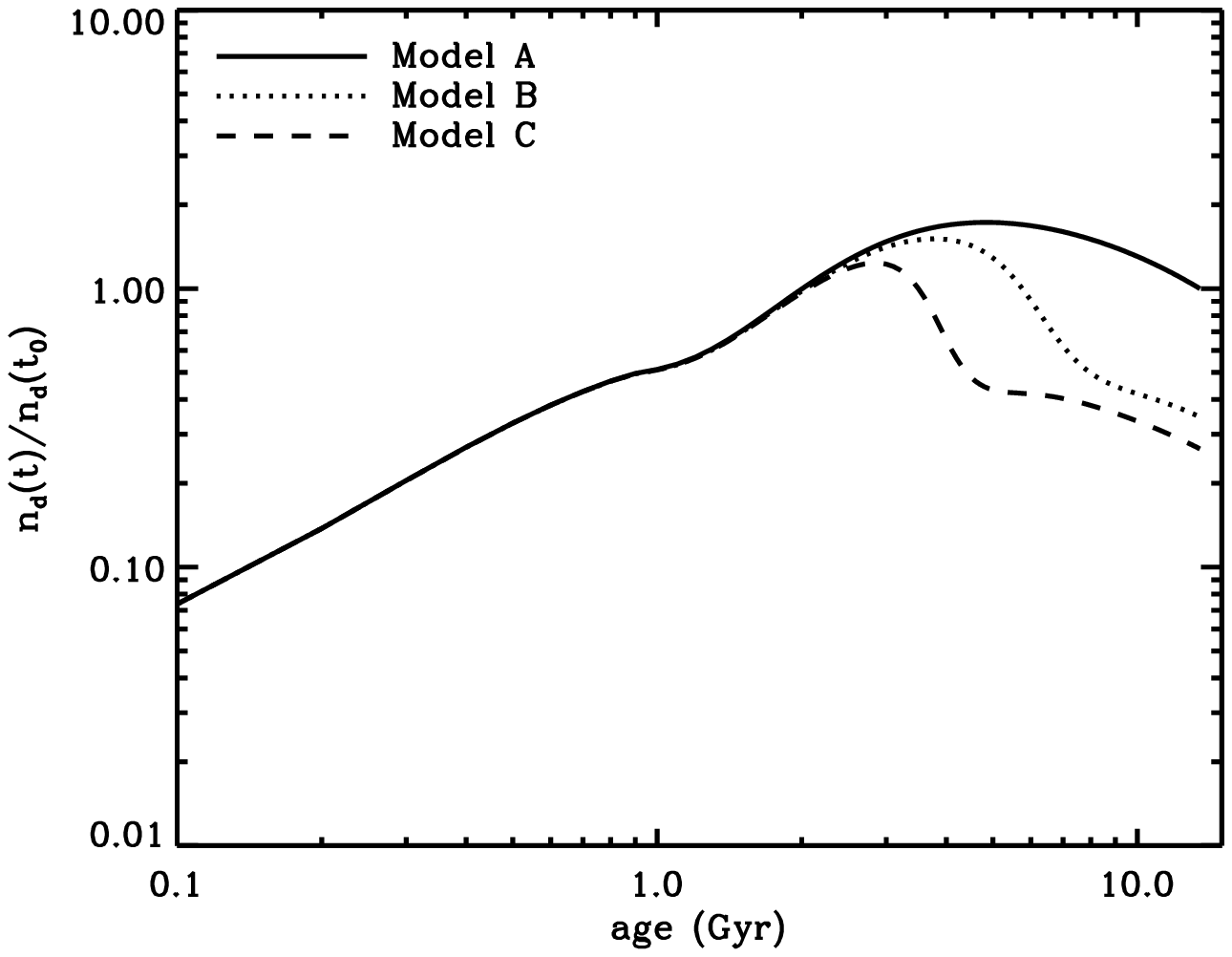}
\includegraphics[clip=true]{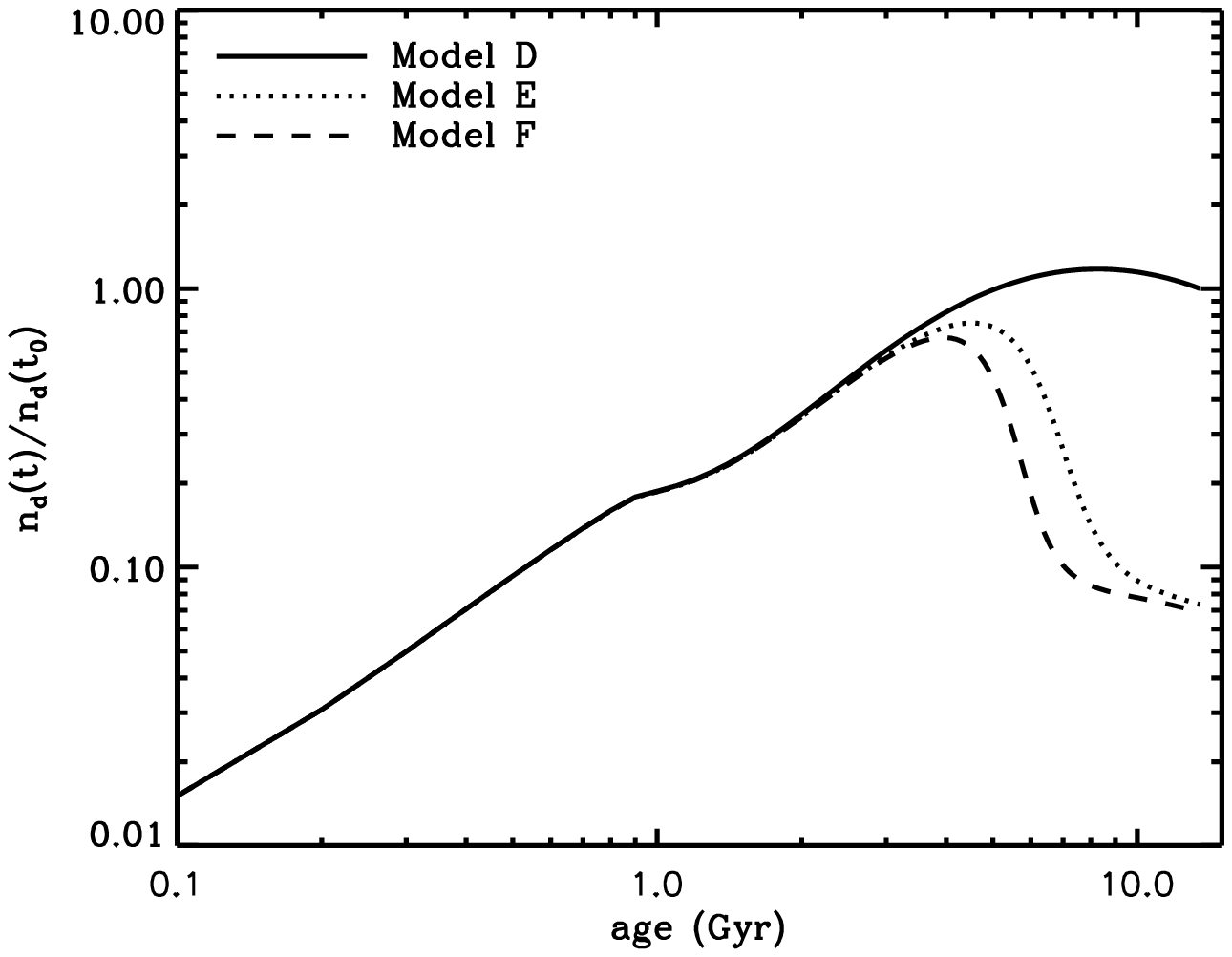}}\\
\resizebox{0.73\hsize}{!}{\includegraphics[clip=true]{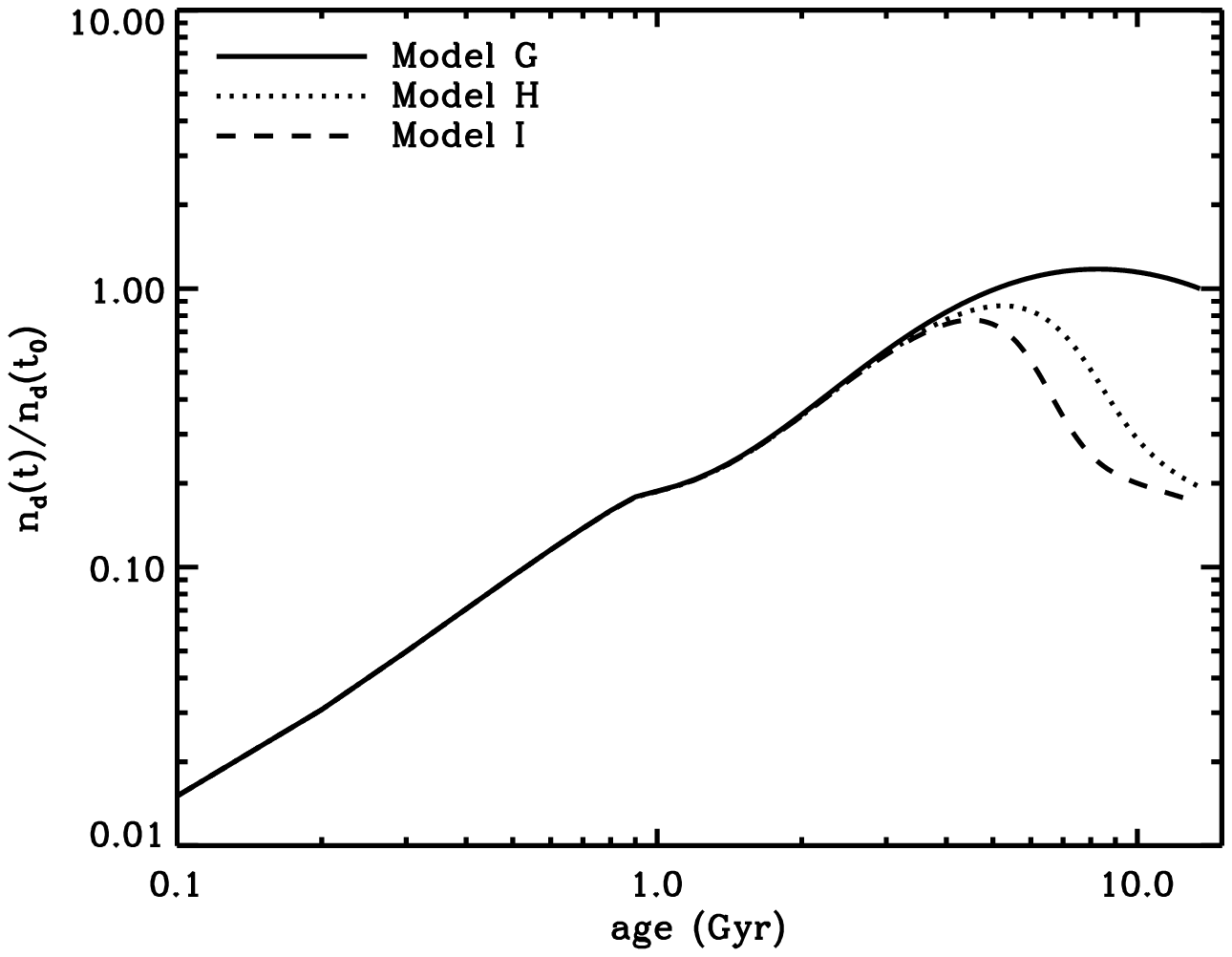}
\includegraphics[clip=true]{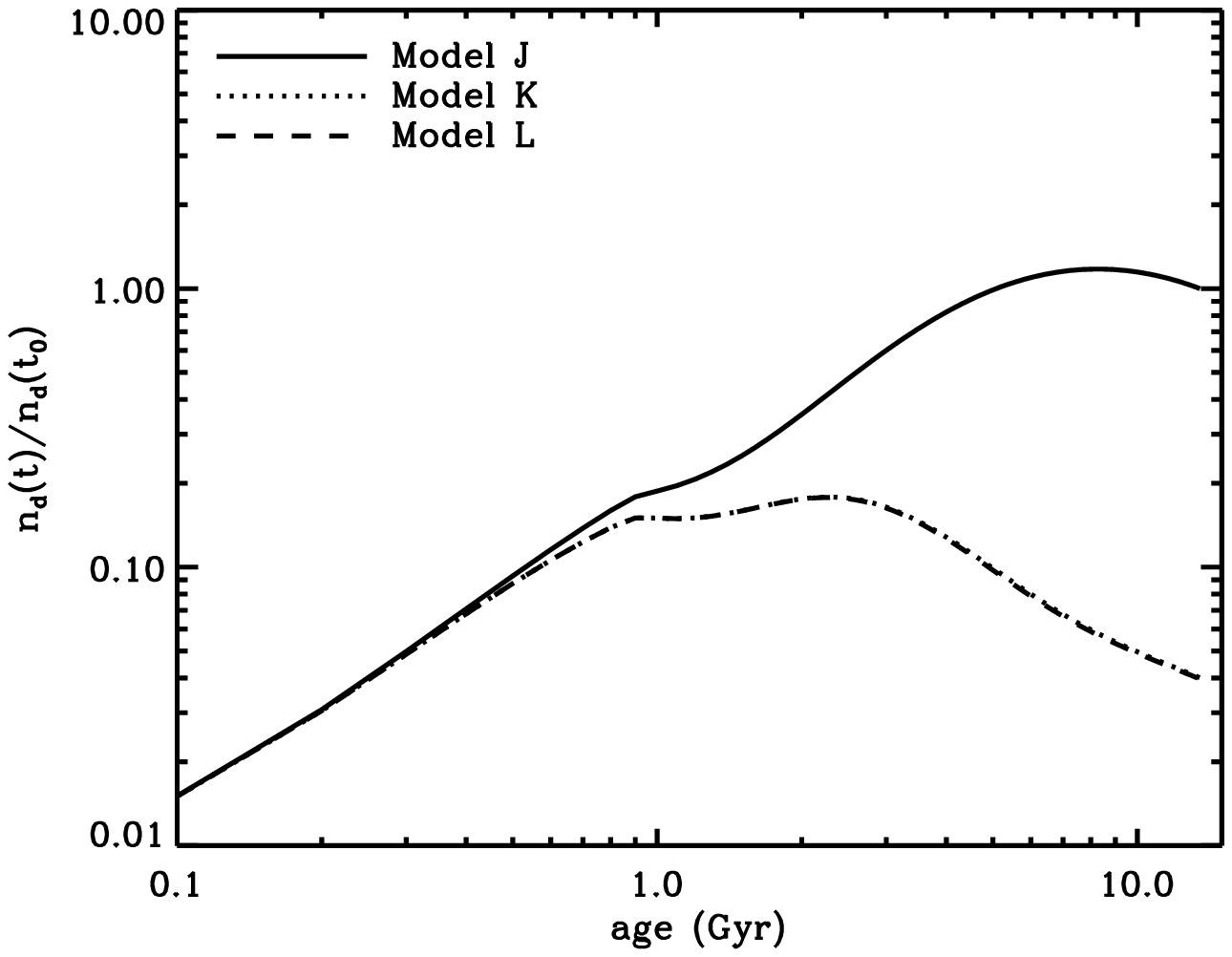}}
\caption{Number density of grains as a function of time normalised to the present-day value in each model. Upper left: the simplistic case ($\upsilon = 0$) of dust destruction due to sputtering associated with the passage of SN shock waves. Upper right: the case of a constant sputtering rate and $\Delta a = 0.001\,\mu$m. Lower left: constant sputtering rate with $\Delta a = 0.001\,\mu$m, but for $a^\star_{\rm min} = 0.05\,\mu$m, i.e., a case where stars inject relatively large grains. Lower right: a case with very substantial ($y_{\rm d} = 0.5\,y_Z$) stellar dust production.
\label{MW_sol_nd}}
\end{figure*}

The evolution of the number density $n_{\rm d}$ of grains is revealing quite a lot about the processing of grains. In the models considered here (which assume no fragmentation), star formation, destructive sputtering and coagulation of grains are processes which are decisive for the number density. The injection of new grains from stars would obviously lead to a monotonous increase of $n_{\rm d}$ if there were no destruction or coagulation. As shown in Fig. \ref{MW_sol_nd}, all models display a peak in $n_{\rm d}$ (normalised to the present-day value for the cases without coagulation: models A, D, G and J, respectively) at some point during the disc-formation phase. Models without coagulation (A, D, G and J) have their maximal $n_{\rm d}$ at the late stages of disc formation ($t_{\rm G} \approx 8$ Gyr, one infall timescale after the onset, at $t_{\rm G} = 1.0$ Gyr), while models with coagulation included display a peak much earlier ($t_{\rm G} \sim 3 - 6$ Gyr). This is a natural result of adding coagulation, since when large enough grains have formed by condensation, coagulation becomes efficient and the rate of small-grain consumption due to coagulation and star formation is greater than the rate of injection of new grains from stars (star formation is rapidly decreasing after a few Gyr, which means the death rate of stars is also falling off quickly). It is interesting, however, to note that suppression of  $n_{\rm d}$ due to coagulation is clearly greater when Galactic dust formation in total is dominated by stellar dust production (cf. models J, K and L, lower right panel in Fig. \ref{MW_sol_nd}). Overall, adding coagulation with an efficiency such that (realistic or not) the coagulation timescale comparable to the condensation timescale, results in a 70--95\% reduction of $n_{\rm d}$ compared to corresponding cases without coagulation (again, see Fig. \ref{MW_sol_nd}). 

With such a large effect on the number density, it is clear that efficient coagulation must have a profound effect on the GSD. Obtaining the GSD from the moments $\mathcal{K}_\ell$ is in practice a nontrivial inverse problem. In theory, however, the problem is straight forward to formulate and solve. Let $W$ be the generating function of the moments $\mathcal{K}_\ell$. $W$ may then be defined in terms of the Laplace transform of $f(a)$, i.e.,
\begin{eqnarray}
\label{generating}
\nonumber
W(s) &\equiv& \{\mathcal{L}(f)\}(s) = \int_0^\infty e^{-sa}\,f(a)\,da \\
& = & \mathcal{K}_0-s\,\mathcal{K}_1+{s^2\over 2} \mathcal{K}_2-{s^3\over 6} \mathcal{K}_3 + \dots,
\end{eqnarray}
where the connection to moments becomes apparent from Taylor expansion of $e^{-sa}$ around $a = 0$. Provided that $W$ is differentiable at $s=0$, the moment $\mathcal{K}_\ell$ is obtained by taking the $\ell$th order derivative of $W$ and evaluate at $s=0$. An approximation of $f(a)$ can be obtained from the inverse Laplace transform of the series on the right-hand side in eq. (\ref{generating}) above,
\begin{equation}
\label{invlaplace}
f(a) \approx \mathcal{L}^{-1}\left[\mathcal{K}_0-s\,\mathcal{K}_1+{s^2\over 2} \mathcal{K}_2-{s^3\over 6} \mathcal{K}_3 + \dots\right].
\end{equation}
This method can be used with Fourier transforms, Melin transforms and possibly other transforms as well. 

\begin{figure*}
\center
\resizebox{0.73\hsize}{!}{\includegraphics[clip=true]{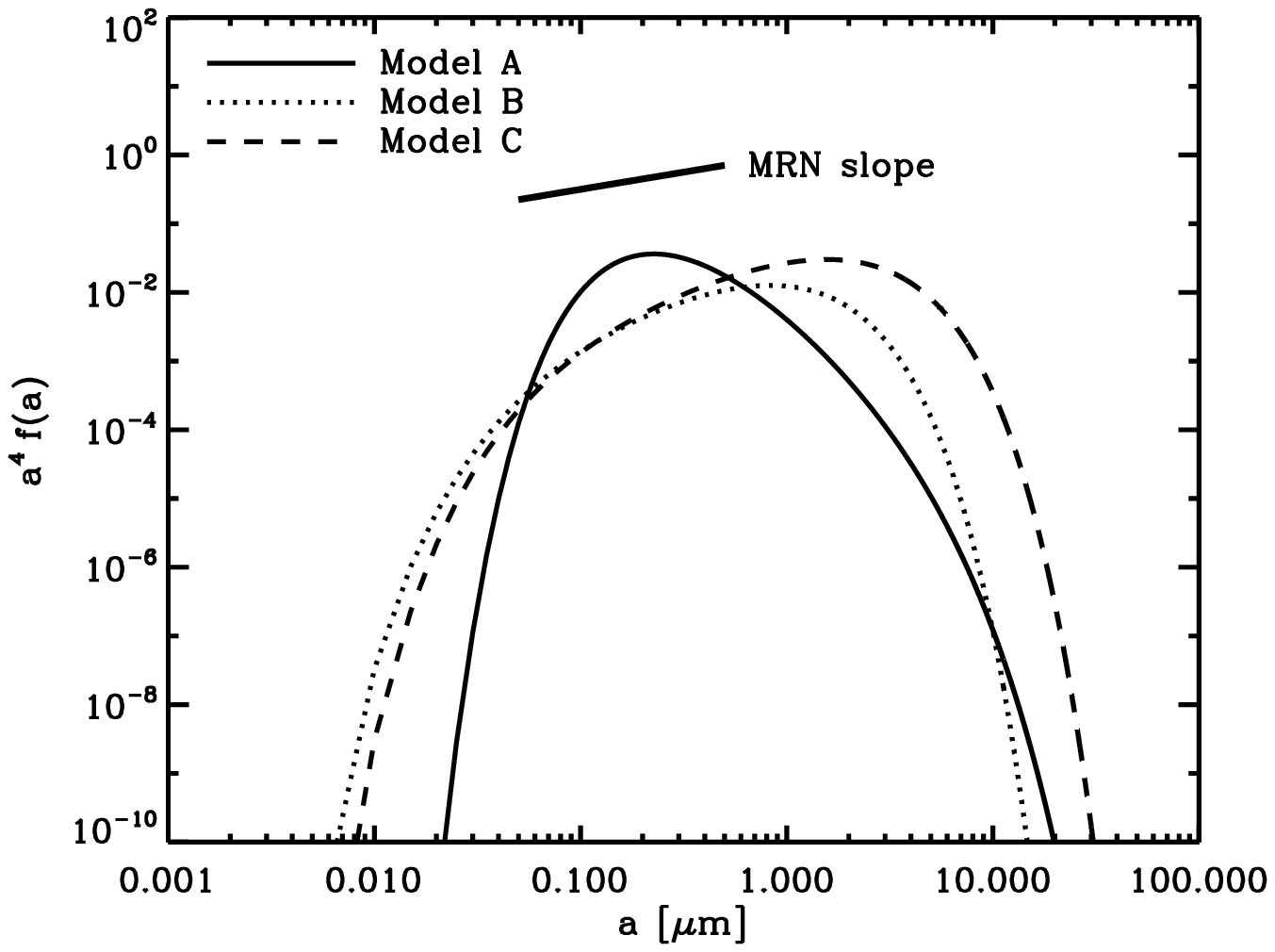}
\includegraphics[clip=true]{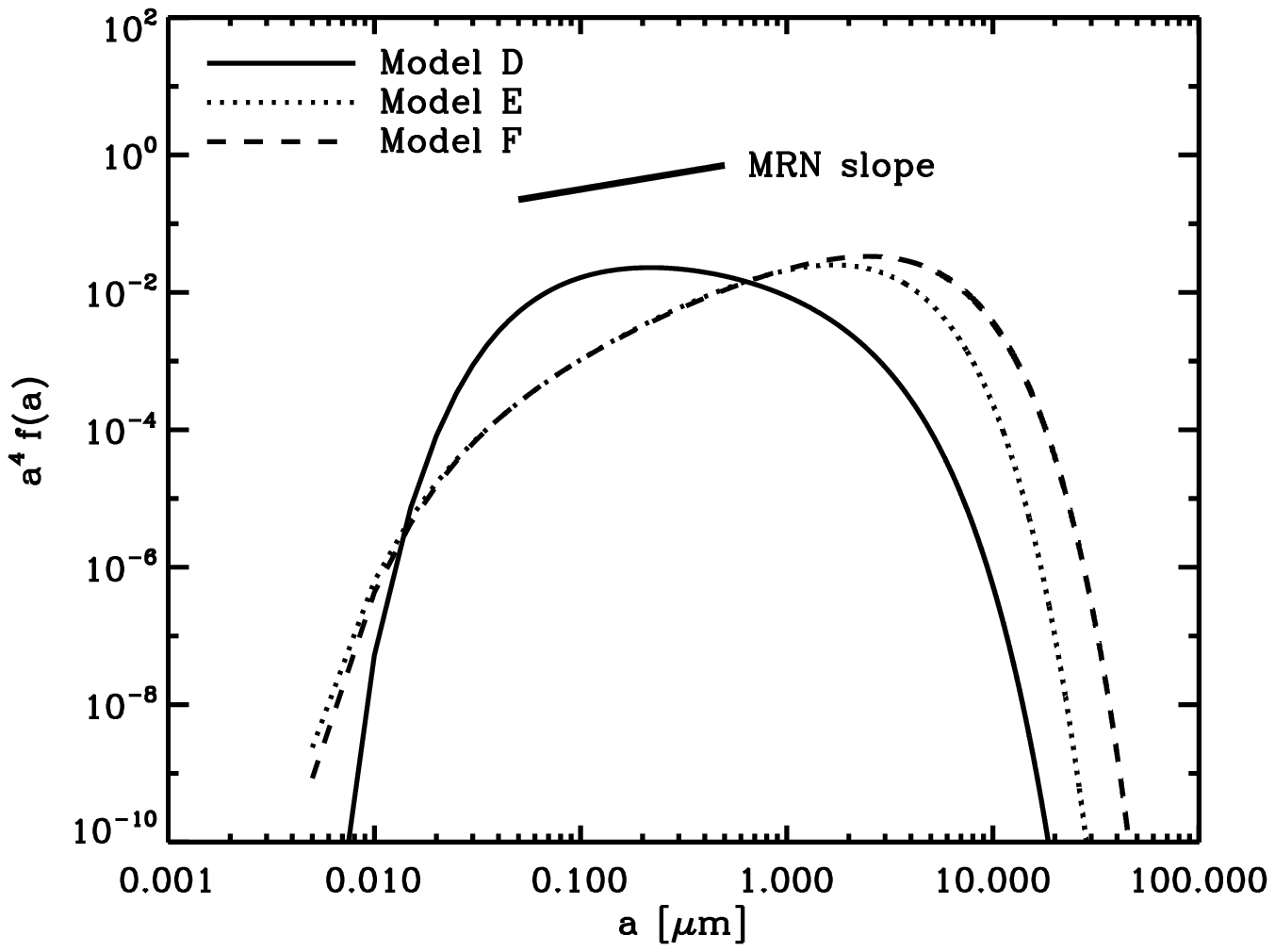}}\\
\resizebox{0.73\hsize}{!}{\includegraphics[clip=true]{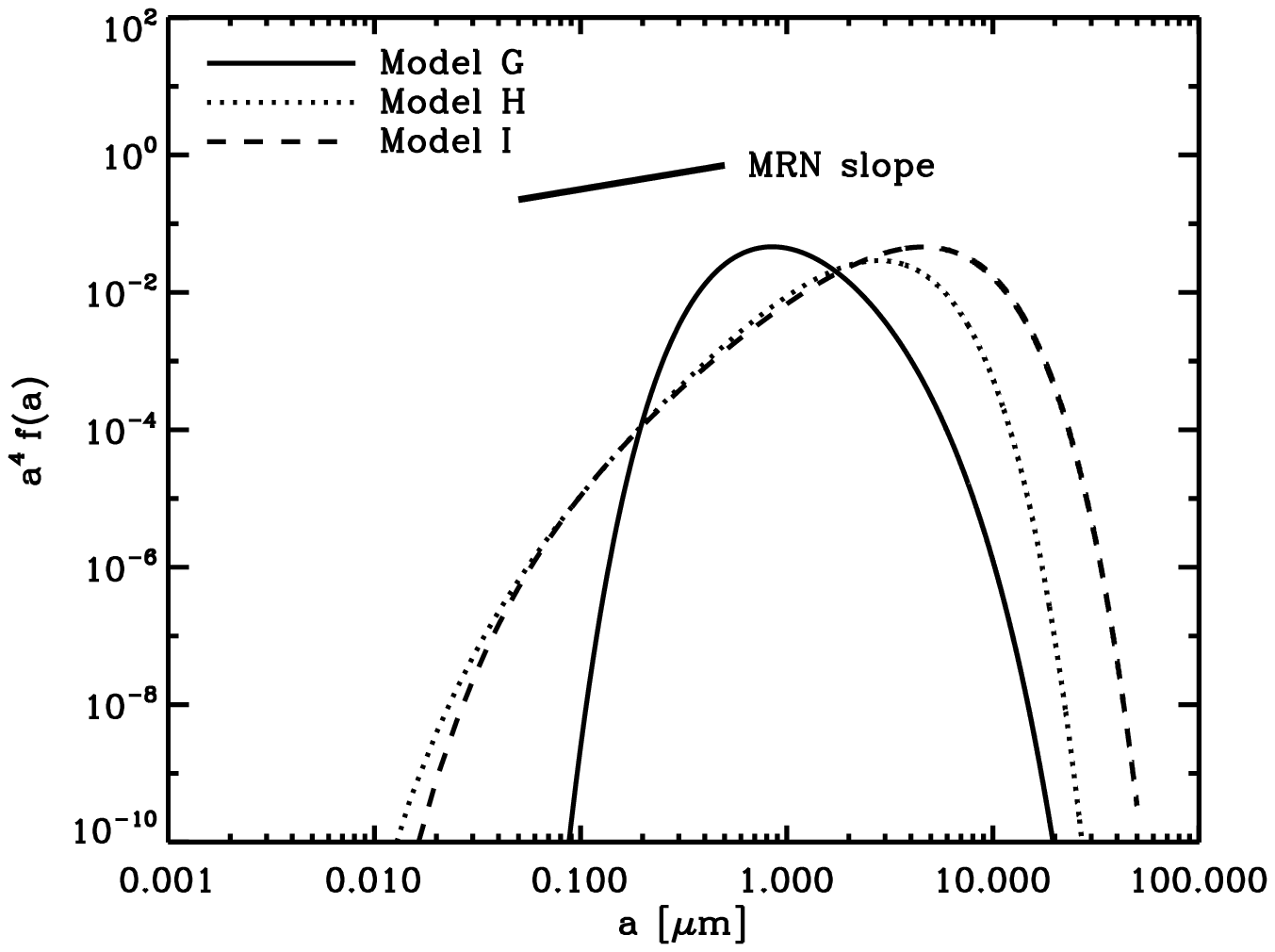}
\includegraphics[clip=true]{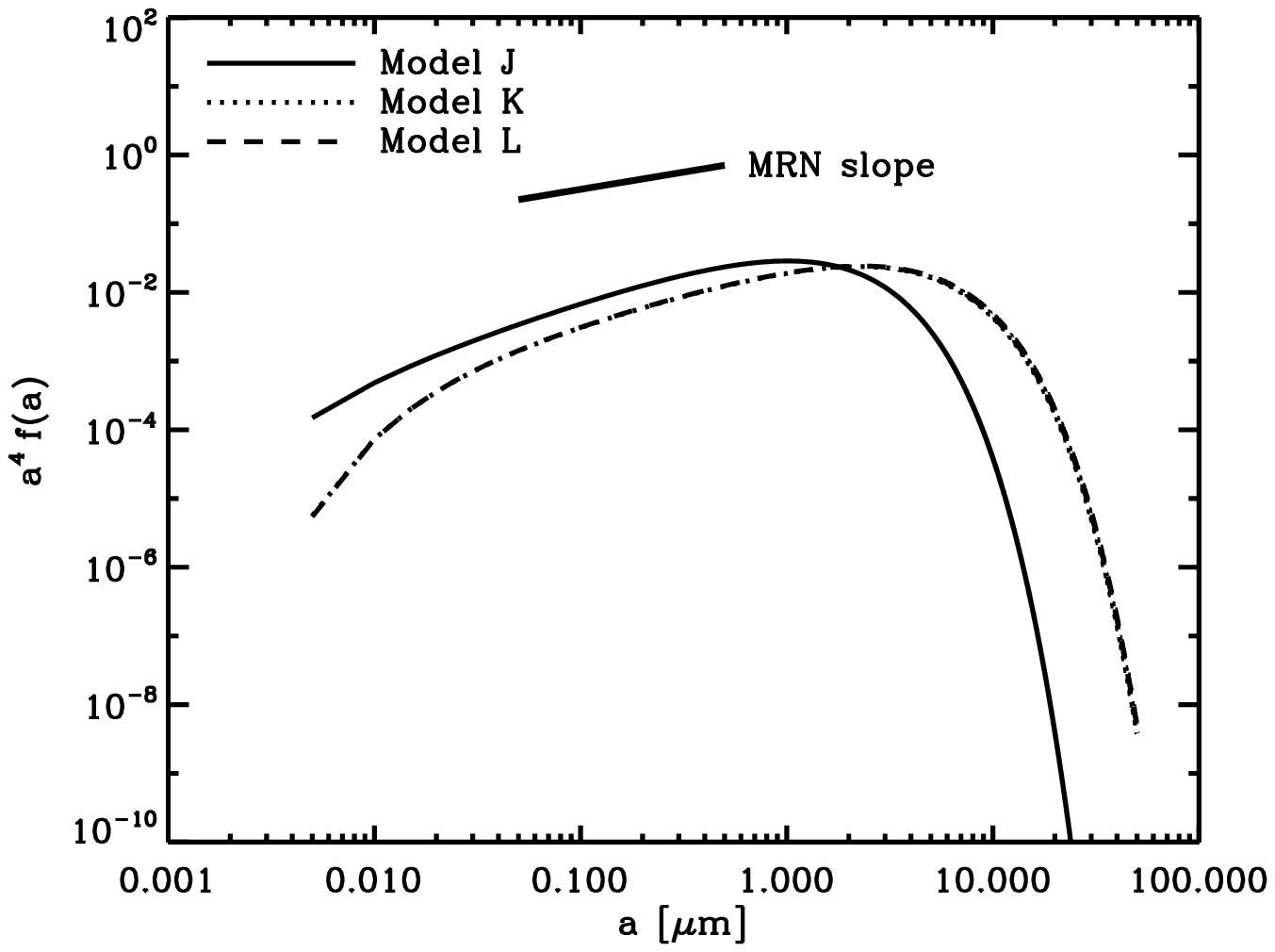}}
\caption{Grain-size distribution at $t=t_0$ (present day), normalised to mass.
 Upper left: the simplistic case ($\upsilon = 0$) of dust destruction due to sputtering associated with the passage of SN shock waves. Upper right: the case of a constant sputtering rate and $\Delta a = 0.001\,\mu$m. Lower left: constant sputtering rate with $\Delta a = 0.001\,\mu$m, but for $a^\star_{\rm min} = 0.05\,\mu$m, i.e., a case where stars inject relatively large grains. Lower right: a case with very substantial ($y_{\rm d} = 0.5\,y_Z$) stellar dust production. Please note that these grain-size distributions are only estimates based on the moment hierarchy and an assumed general functional form of the distributions.
\label{MW_sol_gsd}}
\end{figure*}

\begin{figure*}
\center
\resizebox{0.73\hsize}{!}{\includegraphics[clip=true]{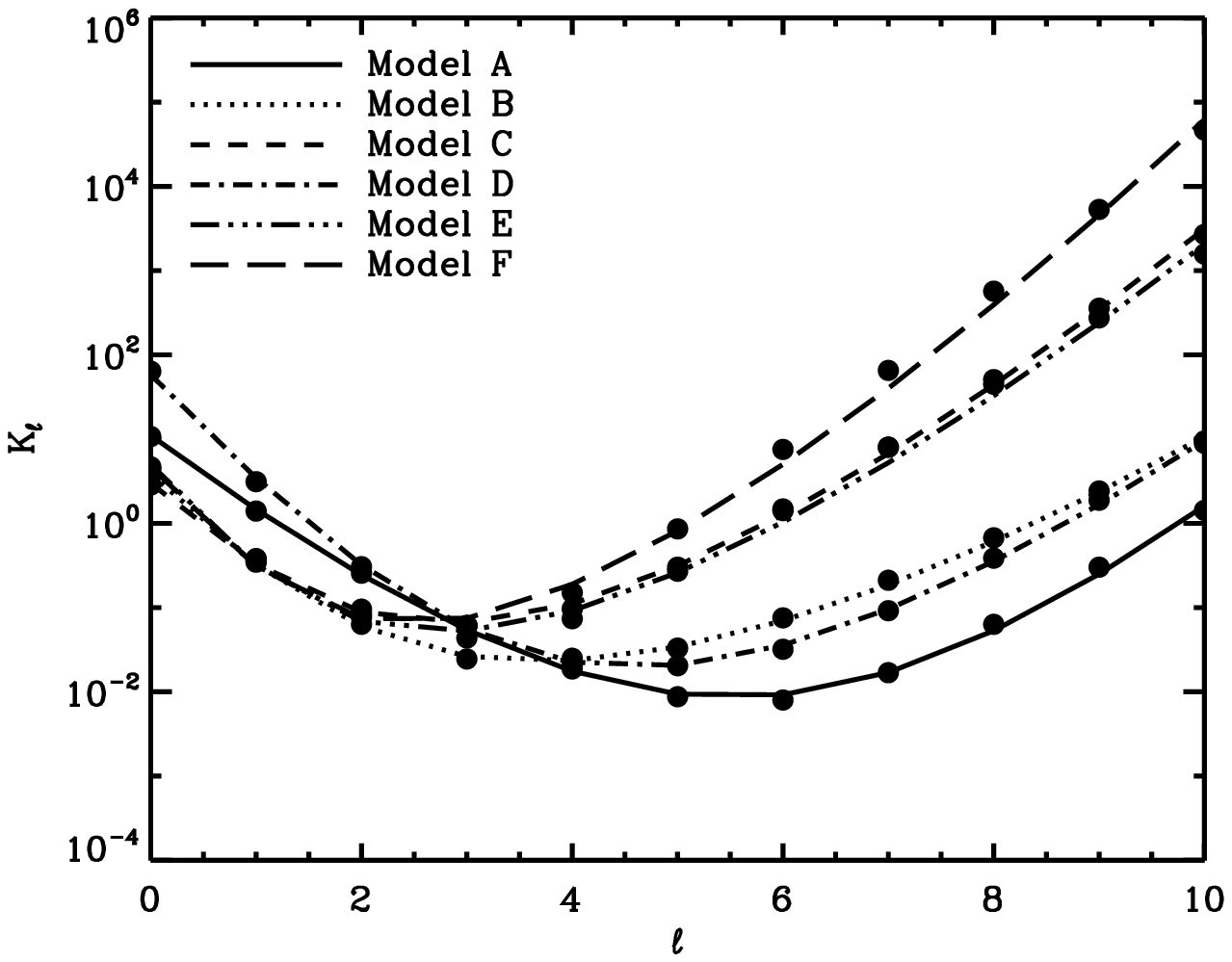}
\includegraphics[clip=true]{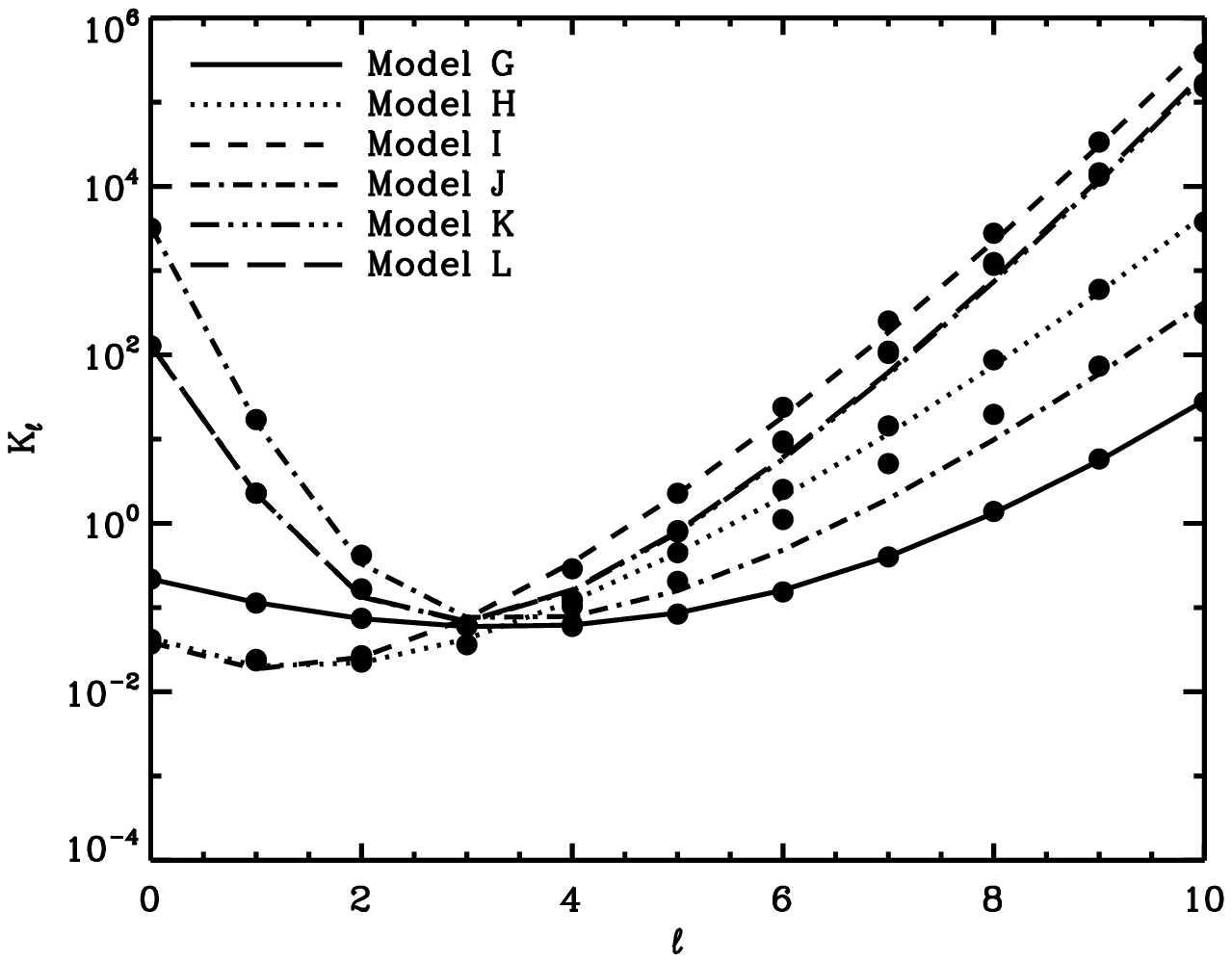}}
\caption{Fits of the moment hierarchy using eq. (\ref{Kfitfun})
\label{MW_sol_fit}}
\end{figure*}

Although transform methods works well in theory, it is a fairly impractical approach as any truncation of the moment hierarchy introduces errors that are difficult to control. However, from eq. (\ref{invlaplace}) it seems that the GSD is determined by its moments. It can in fact be shown that any PDF is uniquely defined by its moment generating function $W$ and, consequently, that a moment hierarchy that extends to infinite order uniquely determines the PDF. Thus, one may safely assume that any functional parameterisation that can yield a moment hierarchy (up to some arbitrary order) which is similar to that obtained from the model is usually a reasonable estimate of the GSD. In the present context, one can make an educated guess about the overall shape of the GSD, which allows use of a more direct and simple method. The GSD of the grains injected by stars is known and defined as a power-law distribution with a negatively defined index $\beta = 3.5$. This suggests there could be a power-law component also in the resultant present-day GSD, especially since coagulation is known to conserve power-law-like GSDs under equilibrium conditions \citep{Dubovskii92}. Coagulation has also a tendency of producing exponential tails, which is easily understood from the fact that the first exact analytical solution to the SCE (for a constant coagulation kernel) found by \citet{Smoluchowski16} is in fact an exponential distribution. At the small-grain end the GSD must turn over and approach zero as grains become arbitrarily small. A reasonable ``guesstimate'' of the  shape of the present-day GSD is therefore obtainable by assuming the functional form
\begin{equation}
f(a) = f_0\, a^{-\beta}\,\exp\left[-\left({a_{\rm to}\over a} +{a\over a_{\rm tail}}\right) \right],
\end{equation}
where $a_{\rm to}$ and $ a_{\rm tail}$ are the turn-over size and the characteristic size where the GSD develops an exponential tail, respectively, and $f_0$ is a normalisation constant. The moments of the above distribution are given by
\begin{equation}
\label{Kfitfun}
\mathcal{K}_\ell = 2f_0\left(a_{\rm tail} \sqrt{{a_{\rm to}\over a_{\rm tail}}}\,\right)^{\ell-\beta+1} K_{\ell-\beta+1}\left(2\sqrt{{a_{\rm to}\over a_{\rm tail}}}\right) 
\end{equation}
where $K_\nu$ is a modified Bessel function of the second kind and order $\nu$. Fitting this expression for the moments  to the resultant moments up to order $\ell = 15$ from the various models (using $f_0$, $\beta$, $a_{\rm to}$ and $ a_{\rm tail}$ as fitting parameters), provides reasonable estimates of the form of the present-day GSD. The fits are generally good (with exception for model J, where the fit to orders $\ell = 6,7,8$ is not satisfactory), which indicates that the ``guesstimate'' is an adequate parameterisation of the GSD. Fig. \ref{MW_sol_gsd} shows the resultant GSDs and the fits to the moments are shown in Fig. \ref{MW_sol_fit}.  

More general, precise and rigorous mathematical methods for deriving a (probability) density function from its moments are possible (Munkhammar, Mattsson \& Ryd\'en, in prep.) and in combination with such inversion methods, also the MOMIC can provide sufficiently detailed information about the GSD for most applications.

\subsection{Net grain-size growth}
\begin{figure*}
\center
\resizebox{0.73\hsize}{!}{\includegraphics[clip=true]{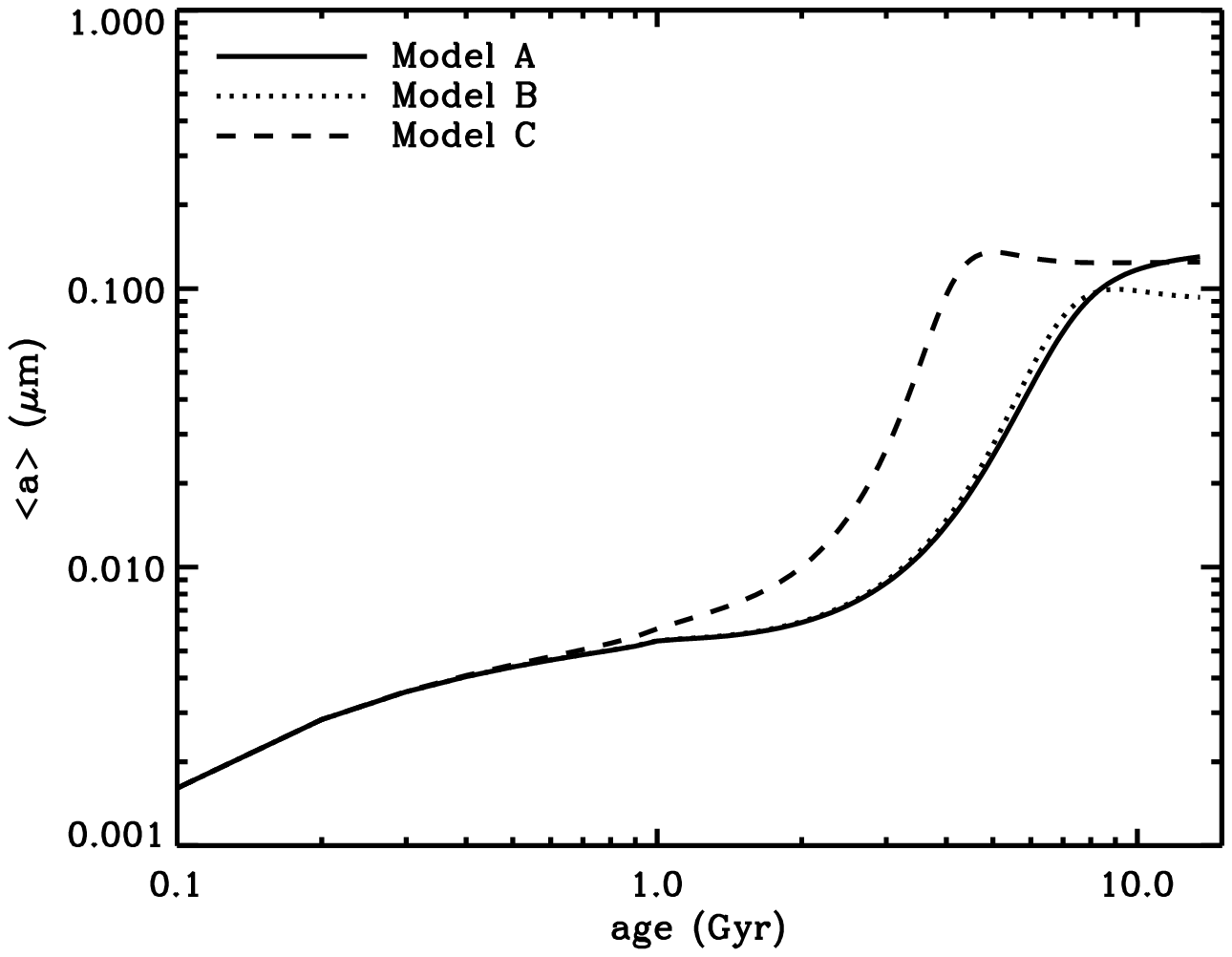}
\includegraphics[clip=true]{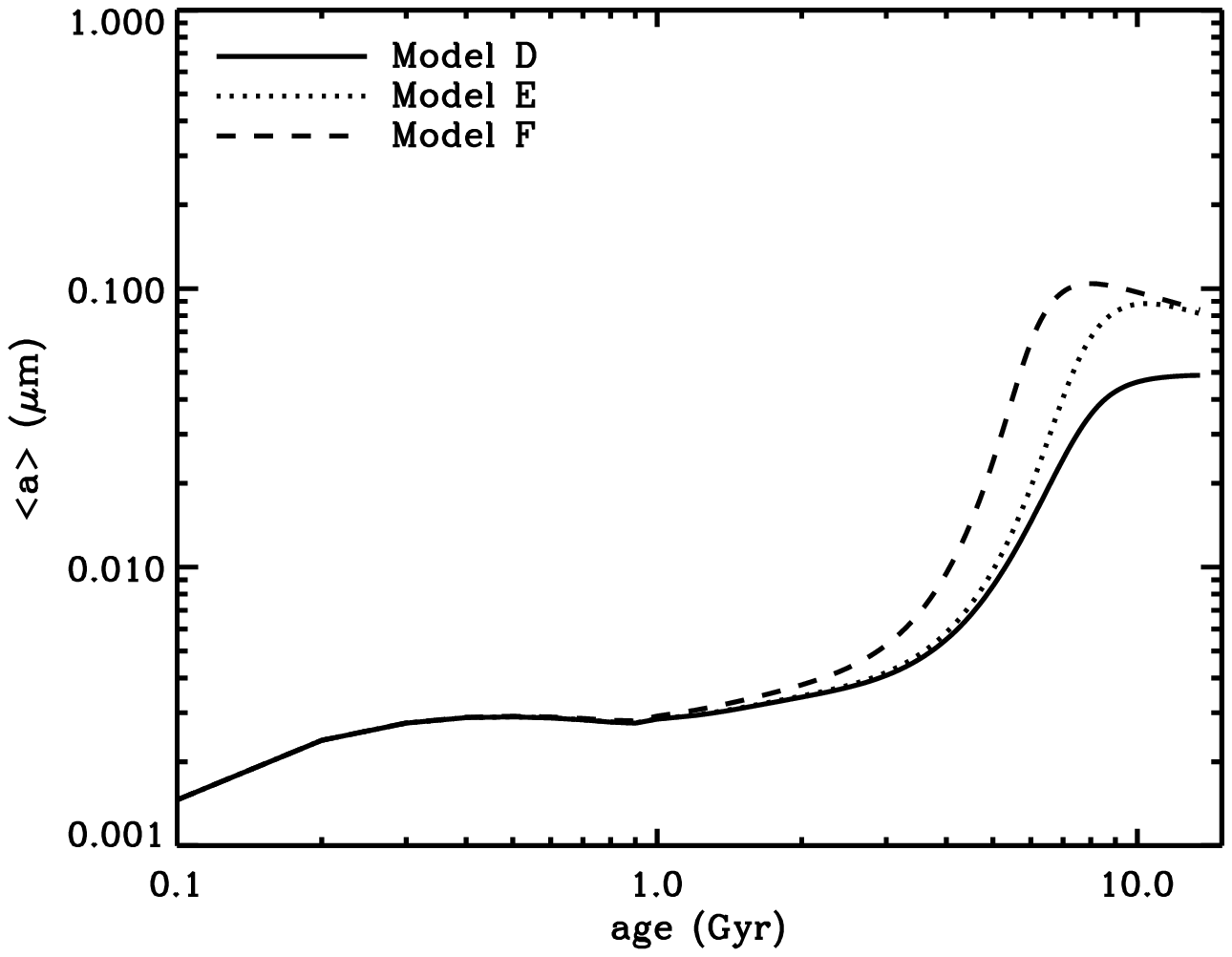}}\\
\resizebox{0.73\hsize}{!}{\includegraphics[clip=true]{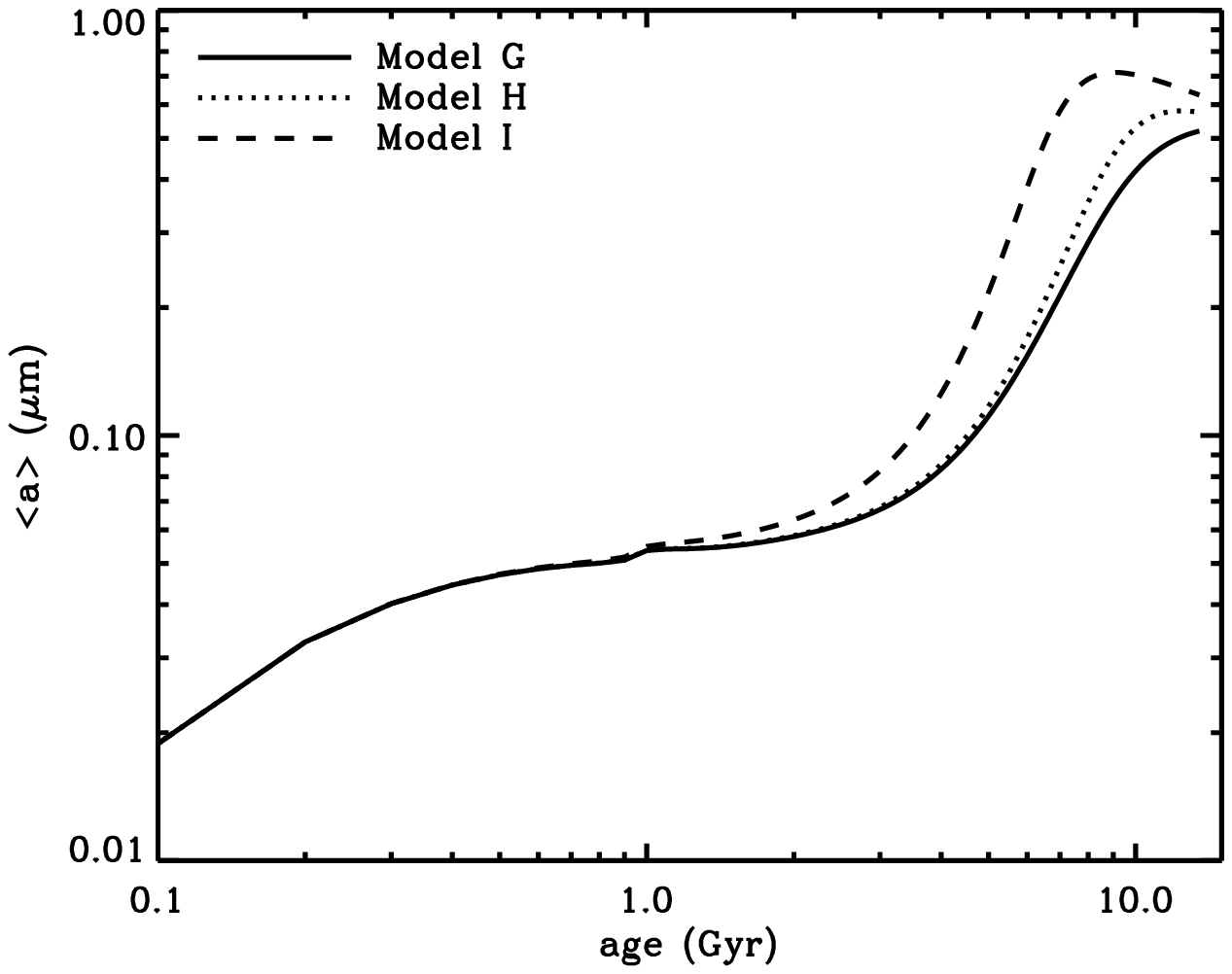}
\includegraphics[clip=true]{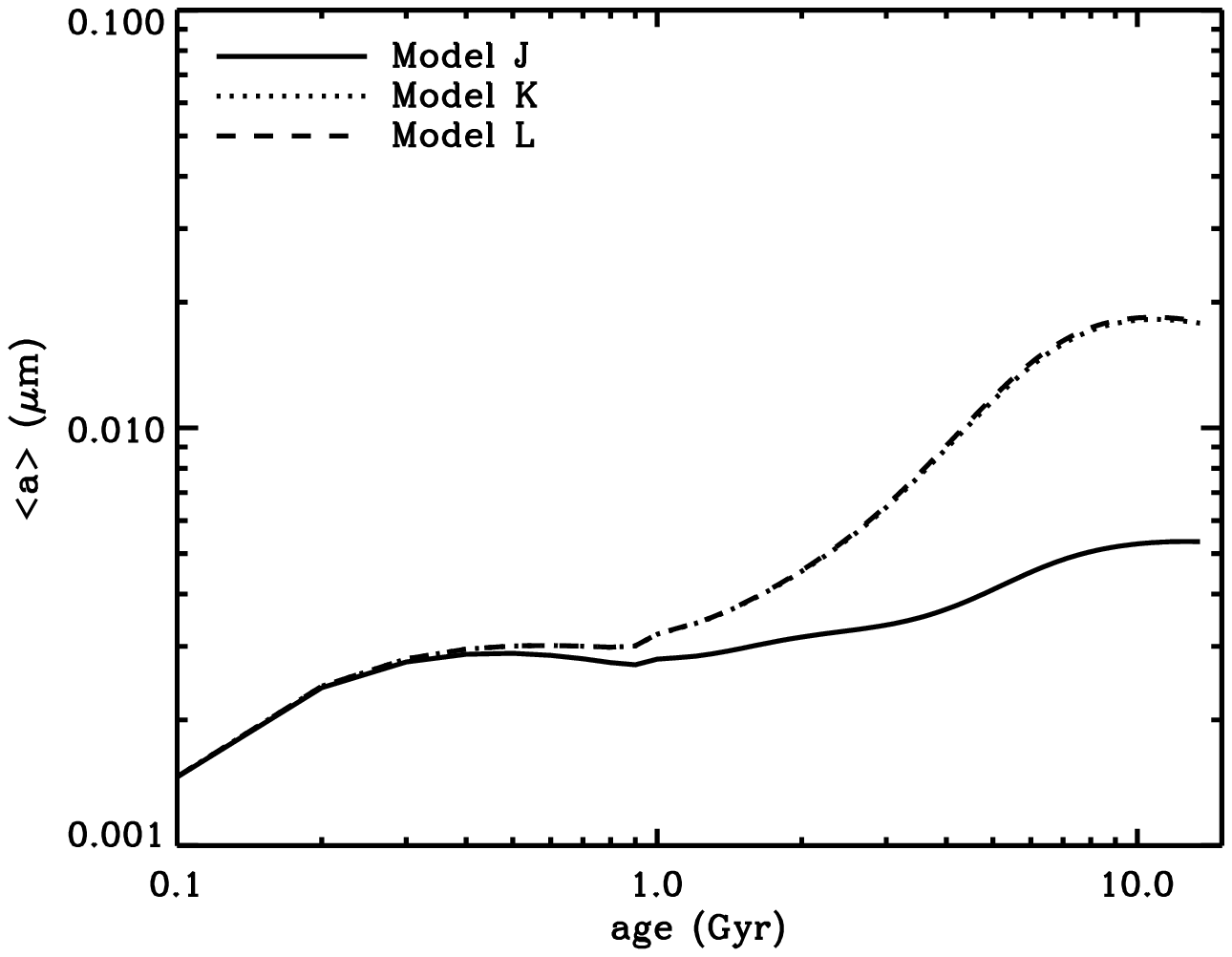}}
\caption{Average grain radius as a function of time. Upper left: the simplistic case ($\upsilon = 0$) of dust destruction due to sputtering associated with the passage of SN shock waves. Upper right: the case of a constant sputtering rate and $\Delta a = 0.001\,\mu$m. Lower left: constant sputtering rate with $\Delta a = 0.001\,\mu$m, but for $a^\star_{\rm min} = 0.05\,\mu$m, i.e., a case where stars inject relatively large grains. Lower right: a case with very substantial ($y_{\rm d} = 0.5\,y_Z$) stellar dust production.
\label{MW_sol_grain}}
\end{figure*}

Most of the grain growth in the ISM is due to condensation in case stellar dust production is moderate. Coagulation makes a contribution only when a sufficient fraction of the grains has reached a sufficient size to make the coagulation rate high enough to affect the average grain size $\langle a \rangle$. This is known as the ``bottleneck effect'', since the transition from the condensation regime to coagulation regime can sometimes be difficult. As shown in Fig. \ref{MW_sol_grain}, the increase in $\langle a \rangle$ is somewhat more pronounced for the cases with size-dependent sputtering (models D to L), which has to do with the shape of the grain-size distribution (more about this below, in Sect. \ref{RBE}). The reason why coagulation cannot be efficient if all grains are below a certain size, is that the most important factor behind efficient coagulation is the total geometric cross section of an interacting pair of grains $\sigma_2 = \pi [a^2 + (a^{\prime})^2]$. When a small grain interacts with a much larger grain, the cross section $\sigma_2$ is dominated by the larger grain, which is why a certain fraction of large grains is needed. Even better, of course, is if all grains are relatively large (but with the same number density), in which case growth by coagulation will be even faster. This effect is clearly seen in the toy models when one assumes that the grains produced by stars are large (about an order of magnitude larger on average compared to the standard case adopted here), since in such case the net increase of the average grain size due to coagulation is a factor of two higher (compare models E and F with models H and I in Fig. \ref{MW_sol_grain}). Note, also, the short  coagulation (reference) timescales $\tau_{\rm coag}^{\rm ref}$ for models G, H and I listed in Table \ref{parameters}.

In case dust production is dominated by stellar dust production (models J, K and L), coagulation is the dominant growth process. Condensation is still as important as a replenishment mechanism, but the average grain size is not much affected. Here, coagulation enters as the dominant growth process as soon as the number density of dust grains is high enough. A sufficiently high number density of smaller grains is the second most important condition for efficient coagulation. Thus, depletion of small grains will halt the growth by coagulation as these are needed as the ``growth species'' in the coagulation process. This means that grain growth, both by condensation and coagulation, becomes inefficient and the average size of the grains actually begins to decrease again (see Fig. \ref{MW_sol_grain}, in particular the upper right panel). This will be referred to as the ``reverse bottleneck effect'' in the discussion below (Sect. \ref{RBE}).

\section{Discussion and conclusions}

\subsection{Dust condensation}
The dust-condensation rate in the ISM is largely controlled by two factors: the availability of the growth species and the total surface area of the grains, where the former is the most important limiting factor. Since the level of dust depletion increases the rate of dust condensation decreases, an effect which can be amplified by the fact that even if the mass of dust is growing (proportional to the third-order moment $\mathcal{K}_3$), the ratio $\mathcal{K}_2/\mathcal{K}_3$ (where $\mathcal{K}_2$ is proportional to the total surface area of the grains) can decrease and further limit the rate of condensation. Even simplistic models of dust condensation, such as the one suggested by \citet{Mattsson12a}, capture the overall evolution of dust depletion due to condensation rather well, but solving the moment hierarchy, and thereby taking the evolution of grain sizes into account, will improve the physical consistency of the model. Using the method of moments, one may also take the interplay between condensation and other types of processing into account. The present paper focuses on the effects of coagulation through implementation of the Smoluchowski coagulation equation (SCE). 

\subsection{Coagulation and large grains}
It is tempting to assume that inclusion of efficient coagulation would automatically lead to the formation of very large grains. But adding coagulation does not immediately lead to a significantly larger average grain radius $\langle a\rangle = \mathcal{K}_1/\mathcal{K}_0$, while the tail of the GSD is extended considerably. The largest grains have radii which are a factor 2--4 larger in the models with coagulation, whereas the average present-day grain size may not increase much at all in some cases (see Fig. \ref{MW_sol_grain}). Very large grains do indeed form, creating an extended tail of the GSD, but why is the average not much affected?

To answer the question above, one has to recall a few facts from previous sections: (1) there is a ``bottleneck'' in the transition from condensation-dominated growth to coagulation-dominated growth; (2) an increase of $\langle a\rangle$ leads to an increase of the volume/area-relation, which in turn yields a longer condensation timescale; (3) the condensation phase ends when the growth species are depleted; and (4) there is a continuous supply of small grains from stars. The ``bottleneck effect'' is a consequence of coagulation being efficient only if there is a sufficient fraction of large grains present, since the overall rate of coagulation is largely determined by the average total cross section of the interacting grains. Thus, growth by condensation is required as a first step. The point when grains become large enough for coagulation to become efficient tend to coincide with the point when the condensation phase ends. For this reason, condensation and coagulation are two processes which are somewhat separated in time. The effect of coagulation on the GSD is mainly shifting the small-grain turn-over towards a slightly larger grain size and extending the large-grain tail. Without a supply of new small grains, the net effect would of course be an increase of the average grain size. But as opposed to condensation, coagulation mainly increases the sizes of already large grains, which are few in number and thus the increase of the average grain radius $\langle a\rangle$ seen in the pure coagulation problem is mostly a result of the reduction of the number density, i.e., the zeroth-order moment $\mathcal{K}_0$. Injection of small grains from stars compensates for the decrease in number density, which in combination with a significant increase of the condensation timescale limits the increase of $\langle a \rangle$. This leads us directly to another phenomenon, which is discussed below.

\subsection{The reverse bottleneck effect}
\label{RBE}
At late stages of ISM processing, a depletion of small grains will halt the growth by coagulation, which means grain growth becomes generally inefficient at some point. From this point on, one can see a decrease in the average grain size, which must somehow be mainly due to production of small grains. In lack of better terminology, this phenomenon will here be referred to as the ``reverse bottleneck effect'' (RBE) because it is a result of moving out of the coagulation regime as opposed to entering the coagulation regime after an initial phase dominated by condensation (the ``bottleneck'' is the transition between the condensation- and coagulation-dominated regimes). 

The RBE occurs after a rapid decrease in the number density of grains due to coagulation (see Fig. \ref{MW_sol_nd}, in particular the upper right and lower left panels). The rapid decrease halts when the density of small grains becomes to low, which can be understood from comparison of the the upper right and lower left panels of Fig. \ref{MW_sol_nd}, corresponding to the cases of stellar dust production with and without small grains ($a< 0.05\,\mu$m), respectively. The effect can also be noticed in the estimated GSDs in Fig. \ref{MW_sol_gsd}: the more prominent the RBE seems, the more obvious is the depletion of small grains. This suggests the strength of the RBE depends on the extent of the small-grain depletion during the phase of efficient coagulation.

But why does the average grain size start to decrease? The answer is that when neither condensation, nor coagulation, can sustain a continued growth (due to depletion of molecules and small dust grains), grains can only become smaller or remain the size they are. (The size-dependent sputtering in models D--L will reduce the size of all grains to some extent.) In all models, continued production of dust by stars leads to an increase in the number of small grains, which also extracts a decrease of the average grain size. 

\subsection{Reconstruction of the GSD}
Employing the method-of-moments technique with an interpolative closure scheme (MOMIC) have one obvious drawback: the evolution of the GSD is not followed explicitly. The GSD is a crucial ingredient in models of extinction curves \citep[see, e.g.,][]{Draine07} and dust emission. Thus, finding a reliable way to reconstruct the GSD from a truncated moment hierarchy is of fundamental importance. The method use here (least-squares fitting of an analytical expression to the resultant moment hierarchy) is not able to capture all features of the GSD, but should provide a reasonable estimate of the overall shape, since the moments $K_\ell$ contain quite a lot of information about the general shape of the GSD \citep{encyc_math}. As mentioned previously, they are essentially a unique function of the moment order $\ell$ for any given GSD. Thus, only a function (parameterisation) that is sufficiently close to the true GSD, should be able to yield a set of moments that fit the resultant moments $K_\ell$ from the model. A more general method will be implemented in future work, however.

Taken at face value, the simple reconstructions presented in Fig. \ref{MW_sol_gsd} suggest that, over one Galactic age (assumed to be 13.5 Gyr), interstellar processing of dust grains yields a GSD clearly biased towards large grains -- in particular if coagulation is included. Thus, one may conclude that if the net result of grain--grain interactions is efficient coagulation (remember that fragmentation is not explicitly considered), the GSD clearly becomes top-heavy. The coagulation efficiency assumed in the example models is quite high (possibly unrealistically high) but the effect on the number density and the large-grain tail of the GSD is so strong that even a much lower efficiency will yield a noticeable difference (a high coagulation efficiency was chosen to clearly demonstrate the effect). Calibration of the coagulation (and fragmentation) efficiency can be done using direct numerical simulations of the ISM dynamics. \citet{Hirashita09} used the results by \citet{Yan04}, where solution of the Fokker--Planck equation had been separated from the (magneto-)hydrodynamics, but with today's computing facilities, more exact/direct simulations are feasible. Grain--grain interactions in the ISM are important for the evolution of the interstellar dust component, but still poorly understood.

\subsection{Final remarks and future prospects}
In this paper, first results from a newly developed code based on the MOMIC, i.e., solution of a system of coupled ODEs describing the evolution of the statistical moments of the GSD have been presented. The example models considered here are to be regarded, more or less, as simplistic toy models, following only a single generic dust species and assuming several approximations, but the code is fully capable of handling an arbitrary number of dust species and detailed chemical evolution. However, the code still need further development, since, for instance, coagulation can lead to mixed dust species (such as ``dirty silicates''), which is difficult to deal with from a modelling perspective and not yet included in the code. Moreover, in the present paper, there has been no attempt to model any resultant extinction curve, which is perhaps the most important observation constraint available. But the code can in principle handle an arbitrary number of dust species, each with its own GSD, which in combination with data on absorption and scattering cross sections for the modelled species makes it possible to calculate extinction curves at each time step. Thus, following the evolution of the extinction curve is straight forward and a necessary addition in future attempts to move beyond the simplistic toy models of the present paper. It should also be noted that a lower size limit for star dust, $a_{\rm min}^\star = 0.005\,\mu$m, has been introduced, which effectively led to an overall low-end truncation of $f$. Hence, very small grains ($a\sim 10$~{\AA}) was essentially not considered. However, some recent studies suggest tiny amorphous hydrocarbons may be an important ingredient in the Galactic dust component \citep[see, e.g.,][]{Jones13}, which would require a slightly different model setup.

The MOMIC is a computationally very inexpensive method for modelling the processing of interstellar grains, although the shape of the GSD through the course of evolution cannot be followed explicitly. But the overall shape of the GSD can be deduced from the moment hierarchy and for most astrophysical applications this is in fact sufficient. It should also be noted that more sophisticated inversion methods can be used and one particular type, based on a high-order polynomial approximation of the GSD/PDF, is under development (Munkhammar, Mattsson \& Ryd\'en, in prep.) and will be implemented in the next version of the code. In general, the gain in computational speed overshadows the loss of detailed information about the GSD. It also opens up for applications in which the gas dynamics of the ISM is treated by direct numerical simulations (e.g., radiation-/magneto-hydrodynamic simulations in clouding a poly-disperse dust component) and in detailed models of dust-driven stellar winds, such as those by \citet{Mattsson10a,Mattsson11a,Hofner08} in which grain-growth can be essential. 

Dust is in reality cycled between the different phases of the ISM, in which different types of dust processing may dominate. Thus, a multi-phase model of the ISM will be an important future addition to the code. Even if the time-lag effect of mass transfer from the cold neutral phase (MCs) to the diffuse ISM can be dealt with in a one-phase model \citep[see, e.g.,][]{Zhukovska08}, the effects of the (dynamical) evolution of the different phases cannot be treated adequately. The formation and disruption of MCs is probably one of the most important missing pieces in the current version of the MOMIC code, as presented here.

Finally, understanding the evolution of interstellar dust is important to get better handle on the extinction properties of galaxies in various evolutionary stages \citep[see, e.g.,][]{Hirashita09,Hirashita11,Asano14}, but also, in the greater scheme of all things, to understand when and where it is most probable that terrestrial planets like the Earth form around new-born stars. The availability of condensed material is obviously crucial for the formation of terrestrial planets and could perhaps lend some further support to the hypothetical concept known as a ``Galactic habitable zone'' \citep[see][]{Lineweaver04}, as the fraction of dust in the ISM is expected to be larger in the middle region of a galactic disc. Dust distributions and extinction properties along galaxy discs can also help constrain dust evolution models, and a natural next step would therefore be to follow up on the work by \citet{Mattsson12a,Mattsson12b}, which only provided ``proof of concept'' through simple analytical considerations. 

\section*{Acknowledgements}
The author wishes to thank the anonymous reviewer for his/her meticulous reading of the paper and many valuable comments that have improved the paper. The members of the astrophysics group at Nordita, in particular D. Mitra, A. Brandenburg and X.-Y. Li, are thanked for stimulating discussions on condensation and coagulation and valuable feedback in general. J. D. Munkhammar (Uppsala Univ.) is thanked for sharing his insights on stochastic modelling and integro-differential equations. Nordita is funded by the Nordic Council of Ministers, the Swedish Research Council, and the two host universities, the Royal Institute of Technology (KTH) and Stockholm University.

\section*{References}
 \bibliographystyle{elsarticle-harv} 
\bibliography{refs_dust_els}






\appendix
\onecolumn
\section{Moment equations}
\label{ODEs}
The full set moment equations solved by the MOMIC code in the ``toy model'' is not given explicitly in the text. The right-hand side of each equation can be seen as consisting of a stellar dust-production/consumption term and and another term describing what happens to the dust in the ISM. That is, 
\begin{equation}
{d\mathcal{K}_\ell\over dt} = \left({d\mathcal{K}_\ell\over dt}\right)_{\star} + \left({d\mathcal{K}_\ell\over dt}\right)_{\rm ISM} , \quad \left({d\mathcal{K}_\ell\over dt} \right)_\star  =  {y_{{\rm d},\,{j}} \over \varsigma_\star}{K_\ell^\star\over K_3^\star}\dot{\Sigma}_\star (t).
\end{equation}
All parameters and variables in the stellar dust-production/consumption term are described in Section \ref{moms}. The ISM terms for each moment $\ell$ (up to $\ell = 15$), which has an exact expression, are given by
\begin{eqnarray}
\left({d\mathcal{K}_0\over dt}\right)_{\rm ISM} &=& -{C_0\over a_0^2} [\mathcal{K}_2(t) \mathcal{K}_0(t) +  \mathcal{K}_1^2(t) ],
\end{eqnarray}

\begin{eqnarray}
\left({d\mathcal{K}_3\over dt}\right)_{\rm ISM} &=& 3\,\alpha_{\rm s}\,\langle v_{\rm mol}\rangle \,{\rho_k(t)\over \rho_{{\rm gr},\,k}} \left[\mathcal{D}_{k,\,j}^\infty-\mathcal{D}_{k,\,j}(t) \right]\,{\mathcal{K}_{2}(t)} + 
{\upsilon\,\mathcal{K}_3(t)\over \tau_{\rm d}(t)} \left[1- \mathcal{H}_0 \sum_{k=0}^3 {3!\over k!(3-k)!} {\Delta a^k \mathcal{K}_{3-k}(t)\over \mathcal{K}_3(t)} \right],
\end{eqnarray}

\begin{eqnarray}
\left({d\mathcal{K}_6\over dt}\right)_{\rm ISM} &=& 6\,\alpha_{\rm s}\,\langle v_{\rm mol}\rangle \,{\rho_k(t)\over \rho_{{\rm gr},\,k}} \left[\mathcal{D}_{k,\,j}^\infty-\mathcal{D}_{k,\,j}(t) \right]\,{\mathcal{K}_{5}(t)} + 
{\upsilon\,\mathcal{K}_6(t)\over \tau_{\rm d}(t)} \left[1- \mathcal{H}_0 \sum_{k=0}^6 {6!\over k!(6-k)!} {\Delta a^k \mathcal{K}_{6-k}(t)\over \mathcal{K}_6(t)} \right] + \\
&& {2C_0\over a_0^2} [\mathcal{K}_5(t) \mathcal{K}_3(t) +  \mathcal{K}_4^2(t) ],
\end{eqnarray}

\begin{eqnarray}
\left({d\mathcal{K}_9\over dt}\right)_{\rm ISM} &=& 9\,\alpha_{\rm s}\,\langle v_{\rm mol}\rangle \,{\rho_k(t)\over \rho_{{\rm gr},\,k}} \left[\mathcal{D}_{k,\,j}^\infty-\mathcal{D}_{k,\,j}(t) \right]\,{\mathcal{K}_{8}(t)} + 
{\upsilon\,\mathcal{K}_9(t)\over \tau_{\rm d}(t)} \left[1- \mathcal{H}_0 \sum_{k=0}^9 {9!\over k!(9-k)!} {\Delta a^k \mathcal{K}_{9-k}(t)\over \mathcal{K}_9(t)} \right] + \\
&& {3C_0\over a_0^2} [\mathcal{K}_8(t) \mathcal{K}_3(t) + 2\mathcal{K}_7(t) \mathcal{K}_5(t) + \mathcal{K}_6(t) \mathcal{K}_5(t)],
\end{eqnarray}

\begin{eqnarray}
\left({d\mathcal{K}_{12}\over dt}\right)_{\rm ISM} &=& 12\,\alpha_{\rm s}\,\langle v_{\rm mol}\rangle \,{\rho_k(t)\over \rho_{{\rm gr},\,k}} \left[\mathcal{D}_{k,\,j}^\infty-\mathcal{D}_{k,\,j}(t) \right]\,{\mathcal{K}_{11}(t)} + 
{\upsilon\,\mathcal{K}_{12}(t)\over \tau_{\rm d}(t)} \left[1- \mathcal{H}_0 \sum_{k=0}^{12} {12!\over k!(12-k)!} {\Delta a^k \mathcal{K}_{12-k}(t)\over \mathcal{K}_{12}(t)} \right] + \\ 
&& {2C_0\over a_0^2} [2\mathcal{K}_{11}(t) \mathcal{K}_3(t) + 4\mathcal{K}_{10}(t) \mathcal{K}_4(t) + 2\mathcal{K}_{9}(t) \mathcal{K}_5(t) + 3\mathcal{K}_{8}(t) \mathcal{K}_6(t) + 3\mathcal{K}_7^2(t) ],
\end{eqnarray}

\begin{eqnarray}
\left({d\mathcal{K}_{15}\over dt}\right)_{\rm ISM} &=& 15\,\alpha_{\rm s}\,\langle v_{\rm mol}\rangle \,{\rho_k(t)\over \rho_{{\rm gr},\,k}} \left[\mathcal{D}_{k,\,j}^\infty-\mathcal{D}_{k,\,j}(t) \right]\,{\mathcal{K}_{14}(t)} + 
{\upsilon\,\mathcal{K}_{15}(t)\over \tau_{\rm d}(t)} \left[1- \mathcal{H}_0 \sum_{k=0}^{15} {15!\over k!(15-k)!} {\Delta a^k \mathcal{K}_{15-k}(t)\over \mathcal{K}_{15}(t)} \right] + \\ 
&& {5C_0\over a_0^2} [\mathcal{K}_{14}(t) \mathcal{K}_3(t) + 2\mathcal{K}_{13}(t) \mathcal{K}_4(t) + \mathcal{K}_{12}(t) \mathcal{K}_5(t) + 2\mathcal{K}_{11}(t) \mathcal{K}_6(t) + 4\mathcal{K}_{10}(t) \mathcal{K}_7(t) + 2\mathcal{K}_{9}(t) \mathcal{K}_8(t)],
\end{eqnarray}
where, again, all parameters and variables are described in Section \ref{moms}.

In any galaxy evolution models it natural to assume there is no dust present at $t = 0$. Therefore, the initial condition for the moment hierarchy used here is that all moments $\mathcal{K}_\ell = 0$.

\end{document}